\documentclass[a4paper,11pt]{article}
\pdfoutput=1
\usepackage{jcappub}

\usepackage{amsmath}
\usepackage{bm}
\usepackage{xcolor,color,colortbl}
\usepackage[utf8]{inputenc}
\usepackage{hyperref}
\usepackage{url}
\usepackage{graphicx}
\usepackage{multirow,bigdelim}
\usepackage{amssymb}
\usepackage[amssymb]{SIunits}
\usepackage{multirow}
\usepackage{bbold}
\usepackage{verbatim}
\usepackage{comment}
\usepackage{caption}
\usepackage{physics}
\usepackage{subcaption}
\usepackage{orcidlink}

\usepackage{tikz}
\usetikzlibrary{positioning}

\definecolor{darkgreen}{HTML}{339933}

\usepackage{scalerel,tikz}
\usetikzlibrary{svg.path}
\definecolor{orcidlogocol}{HTML}{A6CE39}
\tikzset{orcidlogo/.pic={
 \fill[orcidlogocol] svg{M256,128c0,70.7-57.3,128-128,128C57.3,256,0,198.7,0,128C0,57.3,57.3,0,128,0C198.7,0,256,57.3,256,128z};
 \fill[white] svg{M86.3,186.2H70.9V79.1h15.4v48.4V186.2z}
 svg{M108.9,79.1h41.6c39.6,0,57,28.3,57,53.6c0,27.5-21.5,53.6-56.8,53.6h-41.8V79.1z M124.3,172.4h24.5c34.9,0,42.9-26.5,42.9-39.7c0-21.5-13.7-39.7-43.7-39.7h-23.7V172.4z}
 svg{M88.7,56.8c0,5.5-4.5,10.1-10.1,10.1c-5.6,0-10.1-4.6-10.1-10.1c0-5.6,4.5-10.1,10.1-10.1C84.2,46.7,88.7,51.3,88.7,56.8z};
}}
\newcommand\orcidicon[1]{\href{https://orcid.org/#1}{\mbox{\scalerel*{
\begin{tikzpicture}[yscale=-1,transform shape]
\pic{orcidlogo};
\end{tikzpicture}
}{|}}}}

\title{The 3D clustering of Lyman Alpha Emitters measured with DESI}

\author[a,b,c]{{H.~Ebina}\orcidlink{0000-0002-1080-0955},}
\author[a,b,c]{{M.~White}\orcidlink{0000-0001-9912-5070},}
\author[c]{{R.~Zhou}\orcidlink{0000-0001-5381-4372},}
\author[d]{{A.~Dey}\orcidlink{0000-0002-4928-4003},}
\author[c]{{D.~Schlegel},}
\author[c]{{J.~Aguilar},}
\author[e]{{S.~Ahlen}\orcidlink{0000-0001-6098-7247},}
\author[f,g]{{D.~Bianchi}\orcidlink{0000-0001-9712-0006},}
\author[h]{{D.~Brooks},}
\author[i,j]{{F.~J.~Castander}\orcidlink{0000-0001-7316-4573},}
\author[c]{{T.~Claybaugh},}
\author[k]{{K.~S.~Dawson}\orcidlink{0000-0002-0553-3805},}
\author[l]{{A.~de la Macorra}\orcidlink{0000-0002-1769-1640},}
\author[h]{{P.~Doel},}
\author[c,m]{{S.~Ferraro}\orcidlink{0000-0003-4992-7854},}
\author[n,o]{{A.~Font-Ribera}\orcidlink{0000-0002-3033-7312},}
\author[p,q]{{J.~E.~Forero-Romero}\orcidlink{0000-0002-2890-3725},}
\author[r]{{Satya~{Gontcho A Gontcho}}\orcidlink{0000-0003-3142-233X},}
\author[s]{{A.~X.~Gonzalez-Morales}\orcidlink{0000-0003-4089-6924},}
\author[t]{{G.~Gutierrez},}
\author[c]{{J.~Guy}\orcidlink{0000-0001-9822-6793},}
\author[u]{{C.~Hahn}\orcidlink{0000-0003-1197-0902},}
\author[v,w]{{H.~K.~Herrera-Alcantar}\orcidlink{0000-0002-9136-9609},}
\author[x]{{M.~Ishak}\orcidlink{0000-0002-6024-466X},}
\author[y]{{D.~Kirkby}\orcidlink{0000-0002-8828-5463},}
\author[c]{{A.~Kremin}\orcidlink{0000-0001-6356-7424},}
\author[h]{{O.~Lahav}\orcidlink{0000-0002-1134-9035},}
\author[c]{{M.~Landriau}\orcidlink{0000-0003-1838-8528},}
\author[z]{{D.~Lang},}
\author[aa]{{L.~Le~Guillou}\orcidlink{0000-0001-7178-8868},}
\author[c]{{M.~E.~Levi}\orcidlink{0000-0003-1887-1018},}
\author[ab,o]{{M.~Manera}\orcidlink{0000-0003-4962-8934},}
\author[ac,ad,ae]{{P.~Martini}\orcidlink{0000-0002-4279-4182},}
\author[d]{{A.~Meisner}\orcidlink{0000-0002-1125-7384},}
\author[n,o]{{R.~Miquel},}
\author[af]{{S.~Nadathur}\orcidlink{0000-0001-9070-3102},}
\author[w,c]{{N.~Palanque-Delabrouille}\orcidlink{0000-0003-3188-784X},}
\author[ag,z,ah]{{W.~J.~Percival}\orcidlink{0000-0002-0644-5727},}
\author[ai]{{F.~Prada}\orcidlink{0000-0001-7145-8674},}
\author[aj]{{I.~P\'erez-R\`afols}\orcidlink{0000-0001-6979-0125},}
\author[ac,ad,ae]{{A.~J.~Ross}\orcidlink{0000-0002-7522-9083},}
\author[ak]{{G.~Rossi},}
\author[al]{{E.~Sanchez}\orcidlink{0000-0002-9646-8198},}
\author[am,an]{{M.~Schubnell},}
\author[ao]{{H.~Seo}\orcidlink{0000-0002-6588-3508},}
\author[an]{{G.~Tarl\'{e}}\orcidlink{0000-0003-1704-0781},}
\author[d]{{B.~A.~Weaver},}

\affiliation{
\noindent \hangindent=.5cm $^{a}${Department of Physics, University of California, Berkeley, CA 94720, USA}

\noindent \hangindent=.5cm $^{b}${Berkeley Center for Cosmological Physics, UC Berkeley, CA 94720, USA}

\noindent \hangindent=.5cm $^{c}${Lawrence Berkeley National Laboratory, 1 Cyclotron Road, Berkeley, CA 94720, USA}

Rest of the affiliations are in Appendix \ref{sec:affiliations}.}

\emailAdd{ebina@berkeley.edu}

\abstract{
We present a clustering analysis of Lyman-$\alpha$ emitters (LAEs) using spectroscopic observations from the Dark Energy Spectroscopic Instrument (DESI) of candidates selected from the Blanco/DECam Intermediate-Band Imaging Survey (IBIS). We measure the two-point correlation function and the power spectrum, including cross-correlations with DESI quasars. Using both analytical and halo occupation distribution (HOD) simulation-based modeling, we find a linear bias of $b \sim 2.31$--$2.62$ for LAEs over the redshift range $2.26 < z < 3.41$. The analytical modeling also provides constraints on the strength of radiative transfer effects, while the HOD analysis characterizes the LAE-halo connection across multiple models. Finally, we quantify the magnitude of non-perturbative clustering effects such as Fingers of God in the LAE population, providing essential input for the accurate modeling of LAE-based cosmological analyses in forthcoming high-redshift surveys such as DESI-II.
}

\begin{document}
\maketitle
\flushbottom

\section{Introduction} \label{sec:introduction}

The current generation of Stage IV spectroscopic surveys have mapped much of the three-dimensional large-scale structure (LSS) of the universe with remarkable precision out to $z \lesssim 2$, delivering sub-percent constraints on the expansion history and the growth of structure \cite{DESI-DR1,DESI-DR2}. 
With rapidly improving detector technologies and survey strategies, the precision cosmology frontier can be advanced to significantly higher redshifts, opening access to new physical regimes: stronger primordial signals from a larger comoving volume, a longer lever arm in cosmic time for growth-rate measurements, and more (quasi-)linear modes that are modelable \cite{Ferraro22}.
DESI-II \cite{Schlegel22} is a proposal for the next step, a successor to the Dark Energy Spectroscopic Instrument (DESI) planned to begin survey operations around 2029. Its high-redshift survey will cover $\sim 3000-5000\,{\rm deg}^2$ at $2 \lesssim z \lesssim 3.5$, and it will serve as a pathfinder to full Stage~V experiments such as Spec-S5 \cite{Beseuner25} that would ultimately push the spectroscopic frontier to $z \lesssim 5$.
At present our primary view of large-scale structure above $z\gtrsim 2$ comes from the Ly$\alpha$ forest \cite{McQuinn16,DESI-DR2-LyA}. This probe is powerful but inherently limited by the sparseness of background quasars, whose number density is low and declines further beyond $z \sim 3$, and continuum fitting and astrophysical contaminants that reduce the number of available modes. In addition, the non-linear nature of density-observable relationship limits growth constraints from Ly$\alpha$ auto-correlation measurements \cite{McDonald00,Seljak12,Cieplak16,Chen21,Ivanov24}. Galaxy tracers instead sample the matter field densely and directly with far higher number densities, offering a powerful, independent probe. DESI-II therefore relies on two new galaxy populations: Lyman alpha emitters (LAEs; \cite{Partridge67,Steidel00}) and Lyman break galaxies (LBGs; \cite{Giavalisco02, Shapley11}).
These can be further complemented \cite{Modi17,Wilson19,Sailer21} by Cosmic Microwave Background (CMB) lensing from concurrent surveys such as the Simons Observatory \cite{SimonsObs} as a third, bias-free tracer of the matter field. The combination of these probes will also allow for multi-tracer analyses, breaking degeneracies inherent in cosmological parameter inference \cite{Ebina24}.

LAEs are identified by their strong Ly$\alpha$ emission at rest-frame $1216$\AA\ and they can be selected photometrically with high efficiency using narrow- or medium-band imaging surveys \cite{Partridge67,Steidel00,Ouchi20}. Selection with medium-band filters ($\Delta\lambda \sim 250$--$300$\AA) provides the wide redshift coverage ($\Delta z \sim 0.2$ per band) required for a 3D LSS analysis in DESI-II, while still isolating the Ly$\alpha$ emission line from adjacent continuum bands \cite{Ebina25,Raichoor25}. 
The Intermediate Band Imaging Survey (IBIS; \cite{IBIS}), using the Dark Energy Camera (DECam) \cite{DECam} on the 4\,m Blanco telescope in Chile, has been specifically designed with DESI-II in mind, employing five adjacent medium bands spanning $4000$--$5300$\AA\ ($2 \lesssim z \lesssim 3.5$). 

The clustering of LAEs is central to their use as cosmological tracers, encoding their relation to the underlying matter distribution and the growth of large-scale structure. Characterizing this clustering, and confirming that it can be modeled reliably, are prerequisites for treating LAEs as robust cosmological probes in DESI-II and future high-redshift surveys. This information is encoded in the multipoles of the redshift-space correlation function. The monopole $\xi_0$ constrains the galaxy bias $b$, while the quadrupole $\xi_2$ captures RSD and thereby the growth rate through $f\sigma_8$, where $f$ is the linear growth rate and $\sigma_8$ the matter clustering amplitude.

Early angular clustering measurements established LAE biases of $b \sim 2$--$5$ at $z = 3$--$6$ \cite{Ouchi08,Gawiser07,Kovac07,Ouchi10}, and projected correlation function analyses confirmed correlation lengths $r_0 \simeq 2$--$5\,h^{-1}$Mpc \cite{Gawiser07,Guaita10,Bielby16}. More recently, angular clustering of ODIN LAEs combined with CMB lensing has constrained bias at $z \simeq 3$--$5$ \cite{White24,Herrera25}, and the companion IBIS paper \cite{Ebina25} performed pseudo-three-dimensional clustering on the DESI-II target selection itself, finding $b \sim 1.8$--$2.5$ at $z \simeq 2.5$--$3$.
None of these measurements, however, can access the RSD quadrupole $\xi_2$, and they recover the monopole $\xi_0$ only with reduced SNR owing to the projection. Both multipoles are needed to test the theoretical models and analysis pipeline for DESI-II.

In this work, we present the first measurement of the three-dimensional redshift-space correlation function multipoles $\xi_0(s)$ and $\xi_2(s)$ for LAEs at $2.2\lesssim z \lesssim 3.5$, using spectroscopic data from DESI and targeting similar LAE selections as the planned DESI-II wide survey. These multipoles yield a direct constraint on the LAE bias and the first constraint on $f\sigma_8$ from this tracer population, together with a stringent test of theoretical clustering models. We interpret both with Halo Occupation Distribution (HOD) models and perturbation theory descriptions of RSD, providing calibrated inputs for DESI-II forecasting, survey optimization, and theoretical modeling.

This paper is organized as follows.
In \S\ref{sec:tracers} we describe the physical and observational properties of LAEs as high-redshift tracers and the theoretical framework for their three-dimensional clustering.
In \S\ref{sec:data} we describe the LAE sample and spectroscopic dataset used in this analysis.
In \S\ref{sec:clustering} we detail the measurement of $\xi_0$ and $\xi_2$ and the modeling of the clustering signal. 
Then in \S\ref{sec:HOD} we present the HOD modeling of the clustering and the inferred halo occupation of LAEs.
We conclude in \S\ref{sec:conclusion}.
Throughout, we assume a flat $\Lambda$CDM cosmology with the best-fit \textit{Planck} parameters \cite{PCP18}, use comoving $h^{-1}$Mpc units unless otherwise stated, and quote magnitudes in the AB system corrected for Galactic extinction using the SFD98 dust maps \cite{SFD}.

\section{Lyman alpha emitters} \label{sec:tracers}

\subsection{Physical properties and observational definition}

Lyman alpha emitters (LAEs) are galaxies characterized by their prominent Ly$\alpha$ emission line at rest-frame wavelength $\lambda_{\rm Ly\alpha} = 1216$\AA, typically with low UV continuum levels. 
The observational criterion most commonly applied is a minimum rest-frame equivalent width (REW) of the Ly$\alpha$ line, with thresholds ranging from REW $\gtrsim 20$\AA\ \cite{Gronwall07,Ouchi08} to REW $\gtrsim 50$\AA\ \cite{Garel15} depending on the survey depth and selection methodology.
Although the designation is primarily observational, LAEs have been physically identified as a population of young ($t \lesssim$ few hundred Myr), low-stellar-mass ($M_\star \sim 10^{8-9}\,M_\odot$), actively star-forming galaxies with star formation rates of SFR $\sim 1$--$10\,M_\odot\,{\rm yr}^{-1}$ and compact morphologies \cite{Gawiser06,Gawiser07,Pirzkal07,Nakajima12,Hagen14,Cowie98,Ouchi20}.
These properties place LAEs at the low-mass end of the galaxy population at their respective redshifts, and make them physically complementary to the more massive, higher-bias LBG population. Note that in the broad sense, LAEs and LBGs are not disjoint populations, as galaxies with Lyman break features can also exhibit Ly$\alpha$ emission and vice versa. However, for the purposes of this work and next-generation cosmology surveys, the LBG selection specifically targets galaxies with a bright continuum, making it distinct from LAEs with faint continuum \cite{Shapley03,Shapley11,Ebina25}.

\subsection{Selection methods}

LAEs are selected photometrically by searching for an excess of flux in a band containing the Ly$\alpha$ emission relative to adjacent bands probing the continuum.
In the past, the most common approach has been to use narrow-band filters ($\Delta\lambda \sim 100$\AA) to isolate the Ly$\alpha$ line at specific redshift windows, yielding high-purity samples but limited redshift coverage and target density \cite{Ouchi20}.
The most recent generation of large-area narrow-band surveys, exemplified by the One-hundred-square-degree DECam Imaging in Narrowbands (ODIN) survey \cite{Lee24}, has extended this approach to wide-fields, discovering more than 100,000 LAE candidates at $z = 2.4$, $3.1$, and $4.5$.

For DESI-II, which plans to observe LAEs over $\sim 3000-5000\,{\rm deg}^2$ at $2.2 \lesssim z \lesssim 3.5$ starting $\sim 2029$, medium-band photometric selection is the primary viable strategy.
Medium-band filters ($\Delta\lambda \sim 250$--$300$\AA) target Ly$\alpha$ peaks in windows of $\Delta z \sim 0.2$ per filter, providing the broad redshift coverage and high target density required for a full-shape 3D power spectrum analysis, while still isolating the Ly$\alpha$ emission line from adjacent continuum bands.
The Intermediate Band Imaging Survey (IBIS; \cite{IBIS}) on the Dark Energy Camera has been specifically designed with DESI-II in mind, employing five adjacent medium bands spanning $4000$--$5300$\AA ($2.2 \lesssim z \lesssim 3.5$) with a depth of $m<25$ in the main (wide-field) survey and $m<25.5$ in select deep fields. 
Spectroscopic validation of the IBIS target selection using DESI ancillary fibers has established the redshift distributions, interloper fractions ($f_{\rm int} \lesssim 15\%$), and basic clustering properties of these samples \cite{Ebina25}, and a similar medium-band approach using Subaru Suprime-Cam has been demonstrated by ref.~\cite{Raichoor25}. 
In this work, we will use a combination of medium-band peak selections and a broadband color cut to improve the target selection of LAEs, selecting LAEs up to a fiber-magnitude of $m_{\rm fiber} \simeq 25$ in the medium bands, making it equivalent to the DESI-II wide survey depth. We will discuss this further in \S\ref{sec:selection}.

\subsection{Clustering properties}

The clustering of LAEs encode key information about their large-scale bias and halo occupation. 
At lowest order in perturbation theory, galaxies trace the matter density field with a linear bias $b$, so that the galaxy overdensity $\delta_g$ is related to the matter overdensity $\delta_m$ by $\delta_g = b\,\delta_m$ on large scales, which implies $\xi_g(r) = b^2\,\xi_m(r)$ in real space. As the bias determines the amplitude of the clustering signal, it is a critical input for forecasting the cosmological constraining power of LAEs in DESI-II and future surveys.
Previous studies have identified a strong dependence of the LAE bias on the selection criteria, particularly the Ly$\alpha$ line flux (or luminosity) limit and the REW threshold \cite{Ouchi20}. A compilation of existing measurements is well described by a fit of the form $b(f_{\rm lim}, z) = A(f_{\rm lim},z)(1+z) + B(f_{\rm lim},z)(1+z)^2$ calibrated against observations from multiple surveys \cite{Ebina24}, but nonetheless the large variance and modeling assumptions (such as the power-law approximation $\xi_{gg}=(r_0/r)^\gamma$ and neglecting RSD effects significant for narrow redshift slices) of the past measurements makes the determination of LAE bias for DESI-II from previous measurements non-trivial. 
Ref.~\cite{Ebina25} has measured the bias of LAEs selected with the same medium-bands as the DESI-II wide survey using projected correlation functions, finding $b \sim 1.8$--$2.5$ at $z \simeq 2.5$--$3$, consistent with the bias fits of ref.~\cite{Ebina24} at the relevant flux limits. However, due to the selection criteria and limitations imposed by the fiber assignment during these DESI observations, they were not able to measure the 3D correlation function $\xi$, limiting their constraining power. 

The clustering strength of LAEs and their large-scale bias are tightly connected to their host halo masses and occupation. 
LAEs are generally found to be hosted by $M_h \sim 10^{10}$--$10^{12}\,M_\odot$ \cite{Ouchi20} halos (a factor of $\sim 5$--$100$ less massive than typical LBG halos \cite{Wilson19}), and HOD analyses suggest that LAEs are predominantly central galaxies occuping distinct halos with a small satellite fraction \cite{Gawiser07,Ouchi10}.  This is consistent with their low stellar masses and compact morphologies. As with the bias, the halo occupation of LAEs is also sensitive to the selection criteria, with brighter LAE samples tending to occupy more massive halos \cite{Ouchi20}. 
Recently the halo occupation distribution of LAEs has been explored by matching angular clustering measurements from narrow- and medium-band surveys to HOD models \cite{White24,Ebina25}. While these analyses have provided important insights into the halo occupation of LAEs and the scale-dependence of the bias, the lack of 3D clustering measurements has limited the ability to make robust distinctions between HOD model instances. 

An important complication specific to LAEs is the impact of the radiative transfer (RT) of Ly$\alpha$ photons through the neutral hydrogen in the circumgalactic and intergalactic medium.
Ref.~\cite{Zheng11} argued that RT can introduce a significant scale-dependent distortion in the LAE clustering signal.
Subsequent work suggested that the magnitude of this effect depends strongly on simulation resolution \cite{Behrens18}. More recent simulation-based work have found that RT can significantly suppress the central galaxy LAE fraction in massive halos ($M_h \gtrsim 10^{12}\,M_\odot$) \cite{Khoraminezhad25}. 
Observational evidence for the impact of H{\sc i} on LAE clustering has been suggested by ref.~\cite{Momose21,Matthee24,Banerjee25}, but with a small sample size.
A concern for DESI-II is that RT modifies the LAE selection anisotropically, as the Ly$\alpha$ escape fraction depends on the local density and the line-of-sight velocity gradient. This will introduce an effective ``fake'' growth rate that mimics the RSD signal and is partially degenerate with $f$ \cite{Zheng11,Ebina24}.
Fisher forecasts modeling this effect at linear order find that, for LAE-only analyses, RT can significantly degrade constraints on $\sigma_8$, while affecting BAO constraints marginally; however, in a full multi-tracer analysis (LAE + LBG + $\kappa$), the impact on $\sigma_8$ and other affected parameters is largely mitigated due to the degeneracy breaking from LBG and lensing \cite{Ebina24}. 
Understanding and quantifying the RT contribution on the actual DESI-II LAE selection is nevertheless essential: even a sub-dominant RT bias in the power spectrum quadrupole could shift the inferred $f\sigma_8$, and the effect must be characterized before the DESI-II analysis pipeline can be trusted.

In this work, we will present the first measurement of correlation function and power spectrum monopole and quadrupole for the DESI-II LAE selection, providing strong constraints of the bias, as well as the first test of the RSD signal and theoretical modeling for this tracer population. We discuss this in more detail in \S\ref{sec:clustering}.

\section{Data} \label{sec:data}

In this section, we present an overview of the LAE sample used in this work. The dataset is primarily drawn from the Intermediate Band Imaging Survey (IBIS) \cite{IBIS}, which provides the majority of the imaging data, and the Dark Energy Spectroscopic Instrument (DESI) \cite{DESI}, which supplies the spectroscopic data. We further supplement the imaging with public data from the Subaru Hyper-Suprime Cam (HSC) Strategic Survey Program (SSP) \cite{HSC-SSP}.

\subsection{Imaging data}

The imaging data used for target selections is mainly from the Intermediate Band Imaging Survey (IBIS; co-PI's: A. Dey \& D. Schlegel, NOAO Proposal \# 2023B-184194, 2025B-479281) on the Dark Energy Camera (DECam) \cite{DECam} at the Blanco telescope in Chile, with assistance from broadband photometry in the HSC-wide survey. 
IBIS is a wide-field, medium-band imaging survey, that is planned to cover $5000\deg^2$ \cite{IBIS}. The observations started in 2024A, with the two deep fields (XMM-LSS and COSMOS, each of $\sim$10 $\deg^2$) completed in 2025A. Each deep field is obtained via 92 dithered exposures in each band, achieving homogeneous depth ($m<25.5$) inside a disk of radius $\sim 1.6^\circ$. For this work, we employ the dataset from the XMM-LSS field for target selections (\texttt{tertiary49} program in DESI, \S\ref{sec:DESI}) and from the COSMOS field for the clustering analysis (\texttt{tertiary54} program).

\begin{figure}
    \centering
    \includegraphics[width=\textwidth]{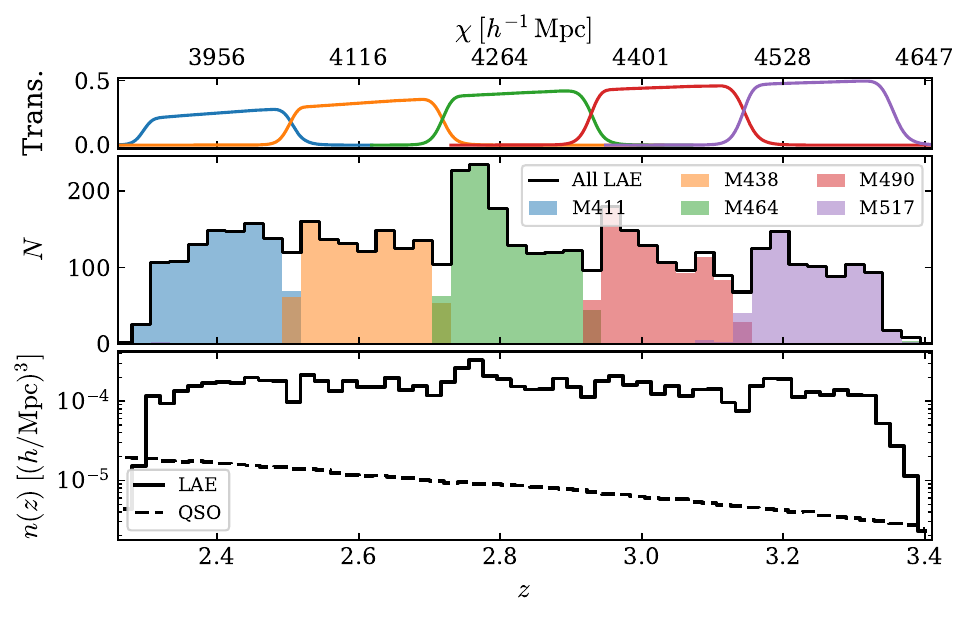}
    \caption{Redshift distribution of the LAE sample selected with the IBIS medium bands. 
    The three panels each show the medium-band transmission curves, the $dN/dz$ of the selected LAEs, and the 3D density $n(z)$ of the LAEs and QSOs. The comoving distance $\chi(z)$ corresponding to the redshift $z$ is shown on the top horizontal axis. 
    }
    \label{fig:dNdz}
\end{figure}

The IBIS filter set comprises 5 contiguous medium bands covering an approximate wavelength range of $4000 < \lambda < 5300$\,\AA, each with a bandwidth of $\sim 260$\,\AA\ (see Figure~\ref{fig:dNdz}). In the deep fields, IBIS delivers medium-band imaging down to a $5\,\sigma$ PSF depth of 25.5 mag. To minimize systematic effects arising from shallower photometric coverage near the field boundaries, we limit the sample to a $1.4\,\deg$ radius in both fields. For the COSMOS field, which is used in the clustering analysis, we additionally exclude the inner $0.3\,\deg$ radius to mitigate systematics caused by reduced fiber completeness near the field center (see \S\ref{sec:DESI} and Figure~\ref{fig:fiber_completeness}). Once regions affected by masked pixels, bad data, and bright-object avoidance are removed, the effective area of the clustering sample is $5.84\,\deg^2$ (Table~\ref{tab:target_summary}).

We supplement the medium-band photometry by cross-matching against the forced-photometry broadband imaging from the HSC-SSP wide survey, which reaches depths of $g=26.5$, $r=26.0$, $i=26.5$, and $z=25.5$ \cite{HSC-SSP}. The broadband data do not enter the core selection, but serve to reject interlopers. Notably, the depths of the relevant HSC wide bands ($gri$) are comparable to those expected from early LSST observations and will therefore be available for DESI-II.

Source detection and photometric measurement are carried out with the \texttt{legacypipe}\footnote{\url{https://github.com/legacysurvey/legacypipe}} \cite{Lang25} and \texttt{Tractor}\footnote{\url{https://github.com/dstndstn/tractor}} \cite{Lang16} framework (see \S8 of ref.~\cite{Dey19}). The pipeline determines source positions and fluxes by forward-modeling the imaging data in localized regions, minimizing $\chi^2$ between model and observed pixels. For each object the pipeline reports a ``total'' flux integrating the full model profile and a ``fiber'' flux predicting the light collected within a $1.5^{\prime\prime}$ diameter assuming $1^{\prime\prime}$ Gaussian seeing. Because the fiber flux closely tracks the signal delivered to a DESI fiber, it is strongly correlated with the spectroscopic SNR and is the quantity entering our brightness cuts (\S\ref{sec:selection}).

\subsection{Target Selection} \label{sec:selection}

In this work we take advantage of the particular filter configuration of IBIS to select LAEs. With five adjacent medium bands, we obtain both a measurement of the Ly$\alpha$ line flux using the ``peak band'' and an estimate of the underlying continuum using the remaining four ``continuum bands.'' All candidates must first satisfy a set of common pre-selection cuts applied uniformly across the five medium bands: the fiber magnitude in the peak band $X$ must lie within the range
\begin{equation}
    22.0 < m_{\mathrm{fib},X} < 25.0,
\end{equation}
together with a blending quality cut $\texttt{fracflux}_X < 1$ and a bright-object veto $m_r > 18$. In addition, we require that the targets pass overall quality cuts. Specifically, we reject objects near bright (\texttt{BRIGHT}) or medium-brightness stars (\texttt{MEDIUM}), or near large resolved galaxies (\texttt{GALAXY}), and objects with saturated (\texttt{SATUR\_X}) or fully masked pixels (\texttt{ALLMASK\_X}) in any of the five IBIS medium bands (M411, M438, M464, M490, M517). 

\begin{figure}
    \centering
    \includegraphics[width=\linewidth]{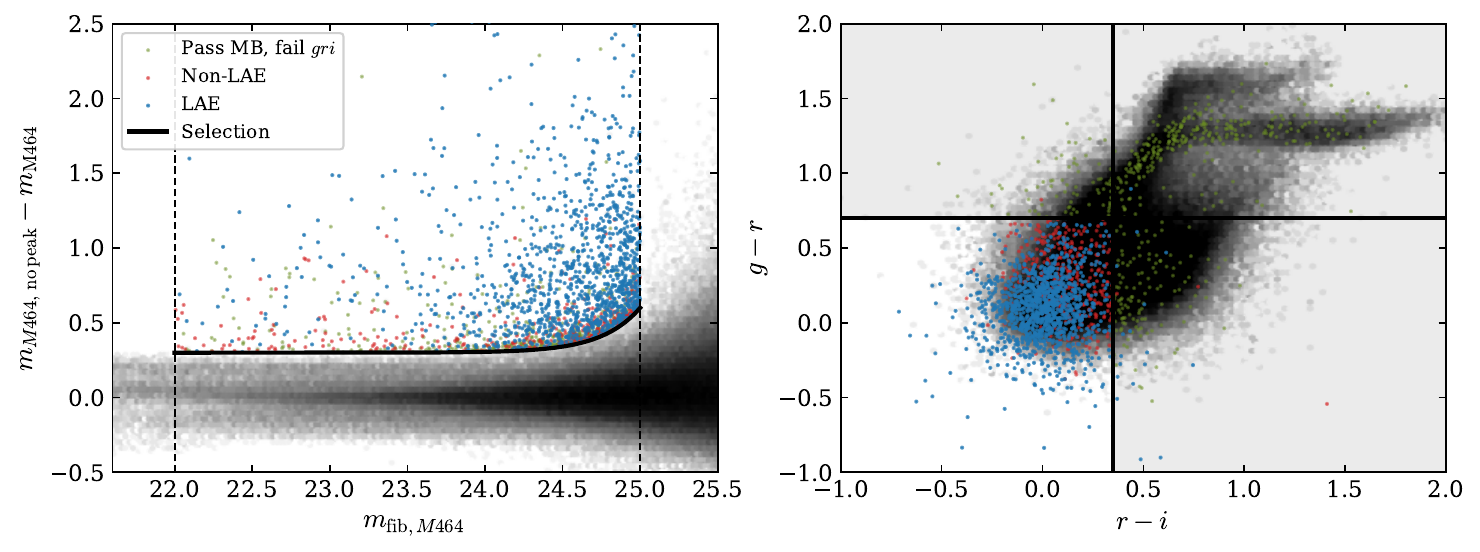}
    \caption{The color selection for LAEs in the \texttt{tertiary54} program as described in \S\ref{sec:selection}, with the medium-band and broadband colors shown on the left and right panels, respectively. 
    Blue and red points are the targeted and observed objects: blue indicates confirmed LAEs, while red indicates the others (interlopers and non-detections). Green points are objects that pass the medium-band color cut (Eq.~\ref{eq:peak_excess}) but are rejected by the broadband cuts (\S\ref{sec:broadband}). Gray points in the background show the parent COSMOS catalog from which the targets are drawn. Note that many green points extend toward the stellar-locus tail in the right panel, indicating that they are likely interlopers.}
    \label{fig:selection}
\end{figure}

The selection criteria are optimized using spectroscopic redshifts from the DESI \texttt{tertiary49} program (in the XMM-LSS field, \S\ref{sec:DESI}) to achieve high purity and target density while maintaining high fiber completeness and a fiber-magnitude limit of $m_{\rm fiber} \simeq 25$ in the medium bands, equivalent to the DESI-II wide survey depth. For clustering purposes, we specifically optimize the selection to minimize the interloper fraction and limit the target density to a level that allows for a high fiber completeness in the follow-up spectroscopy (\texttt{tertiary54} program, in the COSMOS field). The resulting selection criteria are described below and an example for the M464 band is shown in Fig.~\ref{fig:selection}.

\subsubsection{Peak Excess Cut}

For each medium band $X$, we construct a synthetic ``no-peak'' flux as the mean of the extinction-corrected fluxes in the remaining four medium bands:
\begin{equation}
    f_{X,\,\mathrm{no\,peak}} = \frac{1}{4} \sum_{X' \neq X} f_{X'} \ 10^{0.4\, A_{X'}}
    \quad , \quad
    A_{X'} = R_{X'} \ E(B-V)
\label{eq:no_peak_def}
\end{equation}
where $R_{X'}$ is the extinction coefficient appropriate to band $X'$ and the $f_{X'}$ are fluxes in nanomaggies (nMgy\footnote{\url{https://www.legacysurvey.org/dr1/description/}}). The corresponding no-peak magnitude
\begin{equation}
    m_{X,\,\mathrm{no\,peak}}=22.5-2.5\log_{10}(f_{X,\,\mathrm{no\,peak}})
\end{equation}
follows from the standard AB-magnitude relation \cite{Oke83}. 

We select LAE candidates by requiring the peak band to be significantly brighter than the synthetic continuum estimate. To account for increasing photometric noise toward the faint end, we allow the threshold to vary with fiber magnitude:
\begin{equation}\label{eq:peak_excess}
    m_{X,\,\mathrm{no\,peak}} - m_{X,\,\mathrm{peak}} > e^{4\,(m_{\mathrm{fib},X} - 25)}\cdot b_X + c_X,
\end{equation}
where $b_X$ and $c_X$ are band-specific parameters listed in Table~\ref{tab:selection}. At the bright end ($m_{\mathrm{fib},X} \ll 25$) the exponential term vanishes and the cut reduces to the constant floor $c_X$; at the faint limit $m_{\mathrm{fib},X} = 25$ the threshold rises to $b_X + c_X$.

\begin{table}
    \centering
    \caption{Band-specific parameters for the peak excess selection criterion (Eq.~\ref{eq:peak_excess}), $b_X$ and $c_X$, as well as the extinction coefficient $R_X$ (Eq.~\ref{eq:no_peak_def}).}
    \label{tab:selection}
    \begin{tabular}{lcccc}
        \hline
        Band & $c_X$ (floor) & $b_X$ & $b_X+c_X$ (faint end) & $R_X$ \\
        \hline
        M411 & 0.24 & 0.30 & 0.54 & 4.290 \\
        M438 & 0.23 & 0.60 & 0.83 & 4.099 \\
        M464 & 0.30 & 0.30 & 0.60 & 3.877 \\
        M490 & 0.55 & 0.30 & 0.85 & 3.634 \\
        M517 & 0.81 & 0.30 & 1.11 & 3.389 \\
        \hline
    \end{tabular}
\end{table}

\subsubsection{Broadband Color Cuts}
\label{sec:broadband}

The medium-band excess selection is supplemented by broadband color cuts designed to reject low-redshift interlopers. For bands M411 through M490 we require
\begin{equation}
    g - r < 0.7 \quad \text{and} \quad r - i < 0.35,
\end{equation}
while for M517 the cuts are slightly relaxed to
\begin{equation}
    g - r < 0.8 \quad \text{and} \quad r - i < 0.40.
\end{equation}
As an exception, sources with faint enough continuum ($r>25$) are exempt from these color cuts, as they are unlikely to be low-redshift interlopers and the color measurements are noisy. Additionally, if one of the broadbands involved in the color cut is undetected, that color cut is automatically assumed to be satisfied. 

\subsection{DESI spectroscopy} \label{sec:DESI}

\begin{figure}
    \centering
    \includegraphics[width=\textwidth]{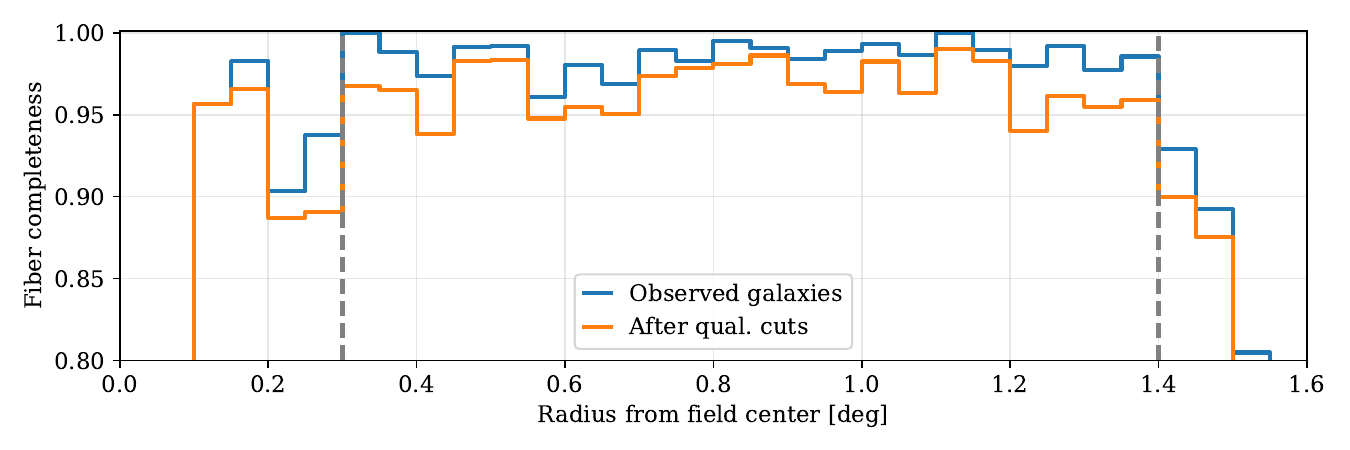}
    \caption{Fiber completeness (the fraction of target galaxies observed) of the LAE sample in \texttt{tertiary54} as a function of angular separation from the field center. The vertical dashed lines at $r=0.3\deg$ and $1.4\deg$ indicate the inner and outer radius cuts applied to the clustering sample to mitigate systematics from lower completeness near the field center. The fiber completeness is $98.6\%$ for $0.3\deg < r < 1.4\deg$, enabling us to analyze the clustering without any fiber assignment correction.}
    \label{fig:fiber_completeness}
\end{figure}

The spectroscopic data are collected using the Dark Energy Spectroscopic Instrument (DESI) \cite{Levi13,DESI}, a Stage-IV survey that operates 5000 robotically positioned fibers over a $3.2\,\deg$ diameter field of view on the 4-m Mayall Telescope at Kitt Peak National Observatory \cite{DESIb,DESI22,Silber23,Miller23,Schlafly23,Poppett24}. Since commencing observations in 2021, DESI has accumulated nearly 50 million extragalactic redshifts through its main cosmology program. DESI also supports ancillary programs that share fiber time with the main survey. We use two such programs, \texttt{tertiary49} ({\sc TileID}s = 83519--83548, in the XMM-LSS field) for target selection optimization and \texttt{tertiary54} ({\sc TileID}s = 83616--83620, in the COSMOS field) for dedicated LAE clustering follow-up. Each DESI pointing covers one tile, assigning targets to the roughly 4000 science fibers (out of 5000 total) based on source positions, focal-plane constraints, and a priority hierarchy. In \texttt{tertiary54}, observations are conducted over five tiles, each with an \texttt{EFFTIME} of one hour, and LAE targets are assigned the highest priority to ensure high fiber completeness. The resulting spectroscopic sample contains 5087 LAEs with secure redshifts. After restricting the clustering sample to a region of $0.3\deg < r < 1.4\deg$ from the field center, the fiber completeness reaches $98.6\%$, enabling us to analyze the target catalog without any fiber-assignment correction schemes. We have validated this choice using mock catalogs subjected to the same target selection and fiber assignment as the data, both before and after the data collection. We find that the clustering signal below fiber-separation scales is affected by $\chi^2\lesssim 0.1$ (Appendix \ref{app:fiberassign}). As shown in Fig.~\ref{fig:fiber_completeness}, additional quality cuts on the spectroscopic observations (faulty fibers and spectrograph depth cuts) remove a further 2\% of galaxies; however, this does not affect the clustering measurement, as it is a small, near-random downsampling.

The spectroscopy is performed using the DESI spectrograph, which covers a wavelength range of 3600--9800\AA\ with a resolution of $R \sim 2000-5200$ \cite{DESI-DR1} and is processed through an extensive spectroscopic reduction pipeline \cite{Guy23}. The redshifts are measured using the DESI pipeline \texttt{Redrock} \cite{Redrock}, which fits templates to the observed spectra to determine the best-fit redshift and classification. 
Specifically for this work, we employ a two-step application of \texttt{Redrock} in order to optimize the redshift measurement for LAEs, which are not included in the standard \texttt{Redrock} template set.
First, we run \texttt{Redrock} in the non-negative matrix factorization (NMF) mode, which fits a linear combination of templates to the spectra without allowing for negative coefficients. In this run we only use LAE template spectra from the visually-inspected DESI spectroscopy over ODIN LAEs \cite{Lee24}, restricting the redshift range to that allowed by the IBIS medium bands. By restricting the fits to the Ly$\alpha$ emission line, this allows us to identify LAE spectra present in the data without increasing sensitivity towards spectroscopic systematics. 
Then, we run \texttt{Redrock} again in the standard mode with the default templates to remove low-redshift interlopers, ruling out high-confidence detections of stars and quasars, as well as objects with two or more emission line detections. 
We confirm the reliability of the redshifts by visually inspecting a subset of the spectra in \texttt{tertiary49}.
This modified redshift fit outperforms a simpler approach of adding high-redshift templates to the standard \texttt{Redrock} run \cite{Ebina25}, finding nearly double the number of secure LAE redshifts.
The resulting redshift distribution of the LAE sample is shown in Figure \ref{fig:dNdz}, where we can see distinct peaks corresponding to the medium bands used for selection.

\begin{figure}
    \centering
    \includegraphics[width=\linewidth]{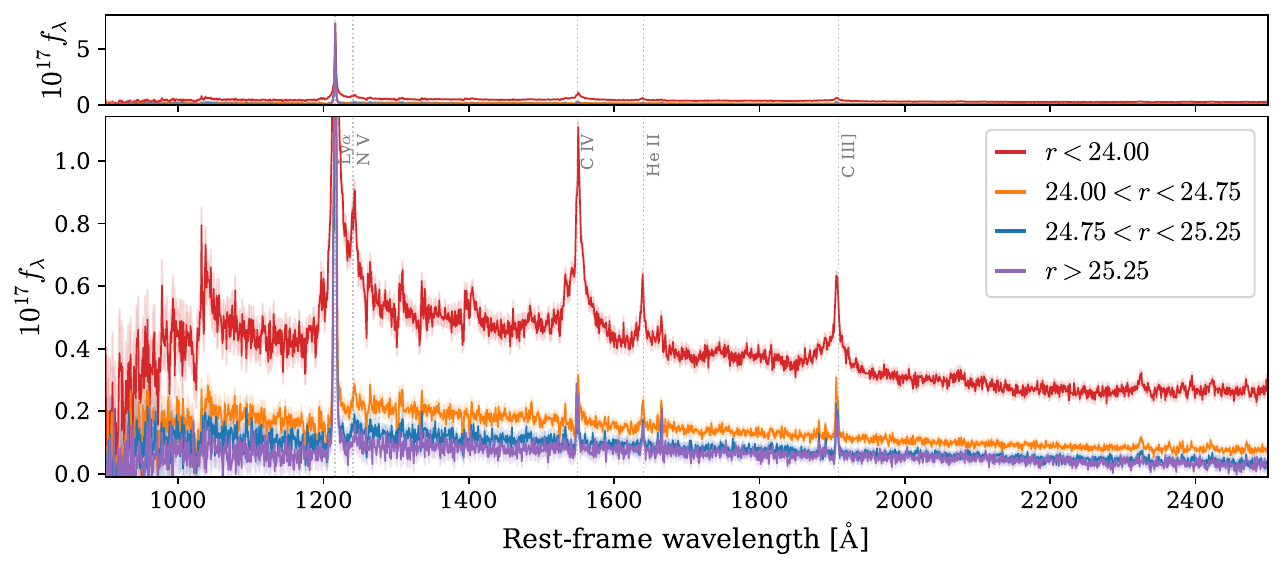}
    \caption{Stacked spectra of the LAEs collected in the \texttt{tertiary54} program.  Spectra are stacked according to the continuum ($r$) magnitude, with the shaded bands showing the propagated $\pm1\sigma$ uncertainty on each stacked spectrum.  The flux is in units of $10^{-17}\,\mathrm{erg}\,\mathrm{s}^{-1}\mathrm{cm}^{-2}\,$\AA${}^{-1}$.}
    \label{fig:stack_spec}
\end{figure}

To characterize the sample and aid its comparison with other observations, in Fig.~\ref{fig:stack_spec} we show composite rest-frame UV spectra of the LAEs, stacked in four bins of continuum ($r$-band) magnitude: $r<24$, $24<r<24.75$, $24.75<r<25.25$, and $r>25.25$. Each spectrum is shifted to its rest frame using the \texttt{Redrock} redshift, $\lambda_{\rm rest}=\lambda_{\rm obs}/(1+z)$, converted to a rest-frame flux density via $f_\lambda^{\rm rest}=(1+z)\,f_\lambda^{\rm obs}$, with the inverse variance transformed consistently as $\mathrm{ivar}^{\rm rest}=\mathrm{ivar}^{\rm obs}/(1+z)^2$, and interpolated onto a common $0.8$\AA\ grid. The composites are then formed as an inverse-variance-weighted mean of these absolute flux densities, where we mask pixels with non-positive inverse variance that indicate rejected data. The composites display the strong Ly$\alpha$ emission characteristic of the sample, together with weaker UV transitions including N\,{\sc v}, C\,{\sc iv}, He\,{\sc ii}, and C\,{\sc iii}], on a faint continuum.

\subsection{Quasars}

In addition to the LAE sample, we also consider the clustering and cross-correlation of quasars (QSOs) in the same redshift range. QSOs are selected using the standard DESI QSO target selection \cite{DESI24-II,Chaussidon23}, which combines optical and infrared photometry to identify quasar candidates. The resulting QSO sample has a high purity and a well-characterized redshift distribution, making it a valuable tracer of the large-scale structure at high redshifts. We use the QSO sample from DESI DR2 \cite{DESI-DR2} for the clustering analysis, which contains $\sim 210$ QSOs at $2.26 < z < 3.41$ over the same footprint as the LAE clustering sample. The QSO cross-correlation with LAEs can provide additional constraints on the bias and RT. 
DESI QSO redshifts are subject to substantial radial-velocity smearing, with $5\lesssim\sigma_v \lesssim 8 \,h^{-1}\mathrm{Mpc}$ and the highest redshift QSOs having the largest smearing \cite{Bault25}. This remains a significant effect even at large scales (at $k = 0.2\,h\,\mathrm{Mpc}^{-1}$, $k\sigma_v \sim 1.2$). This is properly accounted for by the inclusion of counterterms in all fits involving QSOs, as described in \S\ref{sec:analytical_fits}.

\begin{table*}
    \centering
    \caption{Summary of target and observation statistics for the \texttt{tertiary54} field. $N_{\mathrm{targ}}$, $N_{\mathrm{obs}}$, and $N_{z\mathrm{good}}$ are in units of deg$^{-2}$, and $\bar{n}$ is in units of $10^{-4}\,(h^{-1}\mathrm{Mpc})^{-3}$. }
    \label{tab:target_summary}
    \begin{tabular}{llccccccc}
        \hline
        \hline
        Target & Bin & $A\,[\mathrm{deg}^2]$ & $N_{\mathrm{targ}}$ & $N_{\mathrm{obs}}$ & $N_{z\mathrm{good}}$ & $\bar{n}$ & $z_{\mathrm{range}}$ & $z_{\mathrm{eff}}$ \\
        \hline
        \multirow{2}{*}{LAE} & $z\sim2.5$ & \multirow{2}{*}{5.84} & 384.0 & 371.6 & 262.6 & 1.43 & $2.26$-$2.72$ & 2.50 \\
         & $z\sim3.0$ &  & 510.3 & 493.0 & 385.2 & 1.45 & $2.72$-$3.41$ & 2.95 \\
        \hline
        \multirow{2}{*}{QSO} & $z\sim2.5$ & \multirow{2}{*}{10958} & 26.2 & 26.2 & 26.2 & 0.15 & $2.26$-$2.72$ & 2.44 \\
         & $z\sim3.0$ &  & 15.3 & 15.3 & 15.3 & 0.06 & $2.72$-$3.41$ & 2.93 \\
        \hline
    \end{tabular}
\end{table*}

\section{Clustering Measurement} \label{sec:clustering}

The clustering of LAEs has been measured by a number of previous studies, predominantly via the angular correlation function from narrow- and medium-band imaging \cite{Gawiser07,Ouchi10,Guaita10,Bielby16,Khostovan19,White24,Herrera25,Ebina25}. As discussed in \S\ref{sec:tracers}, these measurements rely on modeling assumptions, such as a power-law correlation function and the neglect of RSD in narrow redshift slices, that complicate their interpretation and the comparison across surveys. Here we measure the full 3D redshift-space multipoles $\xi_0$ and $\xi_2$, which directly access the bias and growth rate without these approximations. This is enabled by the high fiber completeness and high-purity selection of our sample, which allow us to measure the 3D redshift-space correlations without any fiber assignment correction. We will measure the correlations both in configuration- and Fourier-space, and interpret the results using both HOD modeling and linear theory with a parametrized RT contribution. In Fourier space, we will also analyze the cross-correlation between LAEs and QSOs, which can provide additional constraints on the bias and RT.

Given the distribution of medium-bands we will analyze the clustering in two redshift bins; the first at $2.26<z<2.72$ (corresponds to M411 and M438) and the second at $2.72<z<3.41$ (corresponds to M464, M490, and M517). We will henceforth refer to the two bins as the $z=2.5$ and $z=3$ samples, respectively.

We generate random catalogs using the \texttt{desitarget} package within DESI \cite{Myers23} and the IBIS DR1 catalogs processed via \texttt{tractor} \cite{IBIS}. Random points are drawn uniformly across the survey footprint at a density of $500,000$ per square degree, corresponding to $\sim500\times$ the density of the data. The brick geometry and pixel-level properties are queried to produce a catalog of random sky positions with associated imaging metadata, including a \texttt{MASKBITS} bitmask encoding pixel-level quality information at each position. We apply quality cuts to the random catalog using the \texttt{MASKBITS} column, mirroring the cuts applied to the data, as described in \S\ref{sec:data}. The randoms are then assigned radial positions (redshifts) using a smooth spline fit to the redshift distribution of data. Due to the small footprint of the field, the choice of spline can induce fluctuations, particularly in the quadrupole. This is discussed further in Appendix \ref{app:nz}.

\subsection{Correlation function}

The correlation functions are computed using the Landy-Szalay estimator \cite{LS}
\begin{equation}
    \xi(s,\mu) = \frac{DD(s,\mu) - 2DR(s,\mu) + RR(s,\mu)}{RR(s,\mu)},
\end{equation}
where $DD$, $DR$, and $RR$ are the number of data-data, data-random, and random-random pairs in the bin defined by the separation $s$ and cosine of the line-of-sight angle $\mu$. The measurements are made using the \texttt{CorrFunc} package\footnote{\url{https://github.com/manodeep/corrfunc}} \cite{Sinha20} over 11 log-spaced bins spanning $0.6 \leq s \leq 20\, h^{-1}\,\mathrm{Mpc}$ and 16 $\mu$-bins. To fit the correlation function, we use covariances derived from pseudo-independent realizations of the mock catalogs, described in \S\ref{sec:HOD}.

\subsection{Power spectrum}
\label{sec:measurement}

We measure the anisotropic power spectrum multipoles $P_\ell(k)$ using the FKP estimator \cite{FKP} implemented in the \textsc{jaxpower} package\footnote{\url{https://github.com/adematti/jax-power}}. 
The density field is painted onto a Fourier mesh using a triangular-shaped cloud (TSC) assignment with third-order interlacing and grid compensation to suppress aliasing. The line-of-sight direction is defined using the ``first-point'' convention appropriate for a wide-angle survey geometry. 

We compute the LAE auto-spectrum, the LAE\,$\times$\,QSO cross-spectrum, and the QSO auto-spectrum, as these three spectra together provide complementary information about the LAE bias and RT contribution. For the LAE auto-spectrum and LAE\,$\times$\,QSO cross-spectrum, we measure the monopole ($\ell=0$), quadrupole ($\ell=2$), hexadecapole ($\ell=4$) and hexacontatetrapole ($\ell=6$), while for the QSO auto-spectrum we only measure the monopole due to the DESI blinding policy. Note that the measurements for $\ell>2$ are not used in the fits in \S\ref{sec:analytical_fits} but rather are used to measure FoG at small scales in \S\ref{sec:FoG}. Furthermore, the QSO quadrupole will not be of significant value in the following analysis, as the QSO bias is already well constrained by the monopole and is negligible in the LAE-QSO cross-spectrum.
The LAE auto-spectrum and LAE\,$\times$\,QSO cross-spectrum are measured jointly from the LAE and QSO catalogs constructed on the DESI \texttt{tertiary54} footprint in the NGC region. The QSO auto-spectrum is measured separately using the DESI DR2 QSO catalog from the \texttt{loa} large-scale structure sample \cite{DESI-DR2}, combining the NGC and SGC Galactic caps.
Given the significantly larger footprint of QSOs over the entire DESI footprint, we will ignore the covariance between the QSO auto-spectrum and other spectra when performing a joint fit later.

The LAE auto-spectrum and cross-spectrum are measured on a mesh with $N_{\rm mesh} = 700$ and cell size $1\,h^{-1}\,\mathrm{Mpc}$, while the QSO auto-spectrum uses a coarser and larger $N_{\rm mesh} = 1152$ and cell size $7.5\,h^{-1}\,\mathrm{Mpc}$ due to the large volume. Power spectra are binned in wavenumber with $\Delta k = 0.02\,h\,\mathrm{Mpc}^{-1}$ over $0.02 \leq k \leq 0.5\,h\,\mathrm{Mpc}^{-1}$. The Poisson shotnoise is not subtracted from the spectra, unless explicitly stated otherwise. 

\begin{figure}
    \centering
    \includegraphics[width=0.98\linewidth]{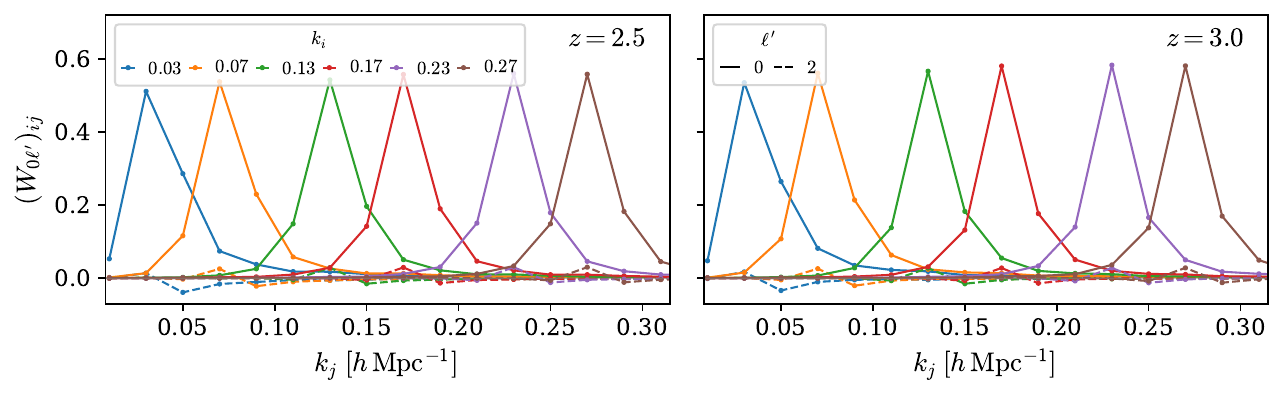}
    \caption{The window matrix $(\mathcal{W}_{\ell \ell'})_{ij}$ with the output fixed at $\ell=0$ and a certain $k_i$ (see legend). The bins are defined with a width of $\Delta k = 0.02\,h\,\mathrm{Mpc}^{-1}$.}
    \label{fig:window}
\end{figure}

For each measured spectrum, we compute the survey window matrix $W_{\ell\ell'}(k, k')$ using the configuration-space method implemented in \textsc{jaxpower} \cite{Castorina18,Beutler21,Chaussidon25}. 
The window function is estimated from the random catalogs using a mesh-based correlation function measurement at two box scales (1$\times$ and 4$\times$ the fiducial box), interpolated onto a logarithmic radial grid via cubic spline, and then transformed to Fourier space using FFTLog. The window matrix maps the ``true'' theory multipoles $\ell' \in \{0, 2, 4\}$ to the observed multipoles. Some slices of the window matrix, with $\ell=0$, are shown in Fig.~\ref{fig:window}.

The analytical covariance matrix is computed under the Gaussian approximation using \textsc{jaxpower}, following the formalism of ref.~\cite{Li19} (Eq.~A.10 for the multi-tracer case). The computation proceeds in two steps. First, the survey window correlation functions are measured from the FKP-weighted random catalogs using a local line-of-sight, yielding the signal-signal (WW), signal-shot-noise (WS), and shot-noise-shot-noise (SS) contributions that encode the mode coupling induced by the survey geometry. Second, these window functions are combined with an input theory power spectrum via a 2D FFTLog transform to produce the full covariance matrix in Fourier space. For the input theory we adopt a Kaiser model evaluated at fiducial parameter values ($b_{\mathrm{LAE}}$, $b_{\mathrm{QSO}}$, and shot-noise $N$ for each spectra). We perform the analysis with varying fiducial values and verify that the final constraints are stable to these changes. The linear growth rate $f(z)$ entering the Kaiser model is computed at the effective redshift of each spectrum.

We validate the Gaussian covariance against the full analytical covariance, including trispectrum and super-sample covariance contributions, computed with \texttt{TheCov}\footnote{\url{https://github.com/cosmodesi/thecov/}} \cite{Wadekar19,Kobayashi23,Alves24} for the LAE auto-spectrum. The diagonal elements are in close agreement and the non-Gaussian, off-diagonal contributions are negligible. We further validate against Monte Carlo covariances derived from mock catalogs in \S\ref{sec:HOD}. The diagonal entries again agree well, while the mock covariance exhibits additional off-diagonal correlations in the monopole, attributable to the Poisson variation in galaxy number across realizations. 

The covariance of the LAE auto-spectrum and LAE$\times$QSO cross-spectrum is computed jointly, preserving their cross-covariance, while the QSO auto-spectrum covariance is computed independently from the separate QSO-only measurement due to the high resolution mesh necessary for the \texttt{tertiary54} footprint. These two blocks are assembled in block-diagonal form for the joint fit, neglecting cross-covariance between the QSO auto-spectrum and the LAE auto/cross-spectrum pair. The cross-covariance neglected here is suppressed by the ratio between the \texttt{tertiary54} clustering footprint ($5.84\deg^2$) and DESI DR2 footprint ($>10{,}000\deg^2$), making this a reasonable approximation.

\section{HOD fits} \label{sec:HOD}

\begin{figure}
    \centering
    \includegraphics[width=\textwidth]{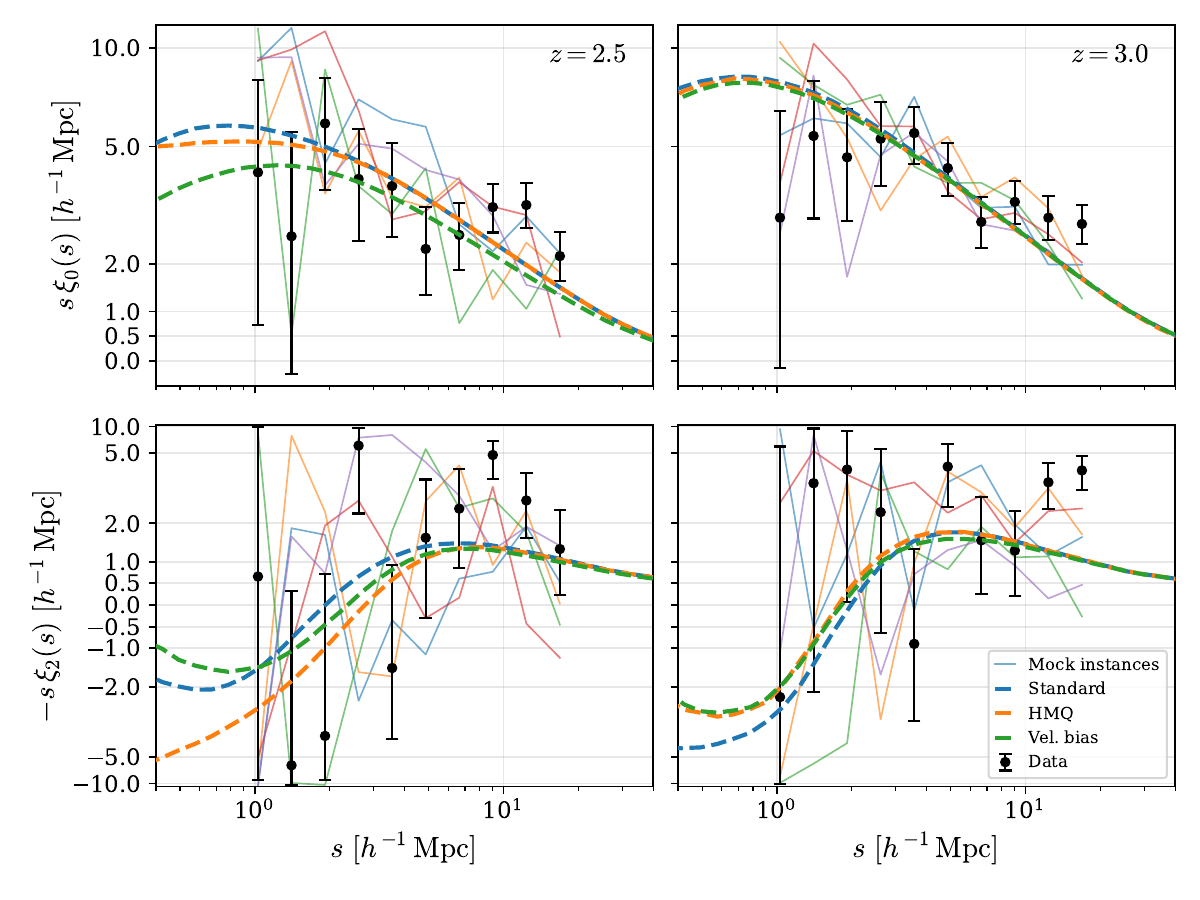}
    \caption{Best-fit HOD models compared to the measured correlation function multipoles for the $z=2.5$ and $z=3$ samples. The data points (black) are shown against three different HOD models (standard HOD in blue, velocity bias in orange, and HMQ in green). The faint solid lines indicate the individual mock sample variation in the correlation function with the standard HOD best-fit model. The errors shown on the data points are from the best-fit model of the standard HOD form.}
    \label{fig:HOD_fit}
\end{figure}

In this section, we use the clustering measurements to model the halo occupation of LAEs within the framework of Halo Occupation Distribution (HOD) models \cite{Wechsler18}. 
An HOD model populates dark matter halos with mock galaxies according to a probability distribution function that depends primarily on halo mass. However, the functional form of the halo occupation for LAEs remains poorly constrained, and there is limited observational evidence to inform how the occupation depends on secondary halo properties beyond mass. 
To address this uncertainty, we fit the data using three distinct HOD models. 
In contrast to the perturbation theory models (\S\ref{sec:analytical_fits}), which benefit from a clear separation of scales in Fourier space, here we work with correlation function measurements, which benefit from a straightforward interpretation in terms of physically distinct distance scales (e.g.~the one-halo regime, the fiber-assignment scale). This approach, for example, enables us to verify that fiber-assignment effects are negligible for this sample, as discussed in Appendix \ref{app:fiberassign}. 
On the other hand, this makes the fit for naturally linear theory (large-scale) parameters, such as linear bias $b_1$, more complicated, partially due to our limited range of scale in a small field. 
We will compress the correlation functions into its monopole $\xi_0$ and quadrupole $\xi_2$, with 10 log-spaced bins from $s_\mathrm{min}=1.0\,h^{-1}\,\mathrm{Mpc}$ to $s_\mathrm{max}=20\,h^{-1}\,\mathrm{Mpc}$ as shown in Fig.~\ref{fig:HOD_fit}. 

While HOD models are originally developed to model galaxies at lower redshifts, both simulation and observational evidence suggest that the functional form\footnote{This may not hold if our galaxy samples contain a mix of different types, or span a wide range of physical properties (e.g.\ stellar mass).  While it is mathematically guaranteed that a perturbative model capable of fitting the clustering of the subpopulations of a mixture will automatically be able to fit the clustering of the mixture itself, this is not true of HOD models.  A trivial example is that a sum of two power laws with different slopes is not a power law.} of galaxy halo occupations are roughly consistent up to $z \sim 3$ \cite{Wechsler18}. Indeed, studies with more complex semi-analytic and hydrodynamical simulations at similar redshifts have shown that existing HOD models are moderately successful at describing LAEs \cite{Nagamine10,Garel15,Gurung-Lopez19,Ravi24,Sullivan25,Khoraminezhad25}. In particular, for medium-band selected LAEs, ref.~\cite{Ebina25} has demonstrated that HOD models are capable of fitting their projected clustering using both standard and Halo Mass Quenching models (\S\ref{sec:standard}, \S\ref{sec:HMQ}). 

For all HOD models, we use the AbacusUtils\footnote{\url{https://abacusutils.readthedocs.io/en/latest/}} software \cite{Yuan22} accompanying the \textsc{AbacusSummit} N-body suite \cite{Maksimova21}, which have been generated using the \textsc{Abacus} N-body software \cite{Garrison21}. Of the available N-body boxes, we use the  high resolution box with a volume of $1\,h^{-3}\,\mathrm{Gpc}^3$ and a particle mass of $m_{\rm part}=3.5\times 10^{8}\,h^{-1}\,M_\odot$, in an attempt to resolve the low-mass halos that host LAEs. 

In order to match the distribution of galaxies in the data, we apply the same angular and radial distribution to the periodic simulation box. The angular footprint of the survey is enforced using a \texttt{healpy} mask with \texttt{nside}$=8192$, constructed using the random catalog. The radial distribution is matched to the spectroscopic redshift distribution from \texttt{tertiary54} by randomly downsampling the input catalog, which also serves to match the number density. As the observational volume of the dataset is significantly smaller than the box volume, randomly offsetting the box before applying this footprint cut can generate 256 pseudo-independent mock catalogs\footnote{Although the factor between observational volume and the volume of the box is smaller than 256, to simulate statistical variations it is not necessary that all realizations are fully independent from each other.} \cite{Ebina25}. We then compute the correlation function multipoles for the mock catalogs using the same pipeline as the data, and fit the HOD parameters by minimizing the $\chi^2$ between the data and the mocks using the covariance inferred from mock catalogs. The model used for the covariance is iteratively changed (within each model) until convergence is attained\footnote{If the iteration becomes cyclical, the model with the least $\Delta\chi^2$ within the cycle will be chosen. This is the case for the standard model at $z=2.5$ and HMQ model at $z=3$.}. Then, using the Akaike Information Criterion (AIC), we choose the preferred model's (the standard model, for both $z$-bins) covariance. To correct for the bias in the inverse covariance matrix estimated from a finite number of mock realizations, we apply the Wishart-Hartlap factor $(N_\mathrm{mocks}-N_\mathrm{data}-2)/(N_\mathrm{mocks}-1)=0.91$ (assuming the underlying data are Gaussian distributed) when computing $\chi^2$ \cite{Hartlap07}.

We note that random downsampling conducted to produce mock catalogs matching the observed radial distribution and number density is clearly an approximation.  The true number density and radial distribution are set by the interplay of the galaxy SED and our selection, which we do not attempt to model.  This has three main consequences.  The first is that in our fits the average halo mass is being constrained by the large-scale bias and the number density plays a much weaker role.  The second is that we do not distinguish between satellites and centrals in our downsampling, while the color-based selection might, and this could alter the halo occupancy.  The third is that by introducing independent random numbers in the sample generation process the stochastic noise becomes much closer to the Poisson value in our mock catalogs than it might be in reality.  A similar effect would be expected in other simulations where scatter between simulated and observable quantities is introduced ``by hand'', e.g.\ the relationship between Ly$\alpha$ REW and SFR in hydro simulations in refs.~\cite{Ravi24,Sullivan25}.  To the extent that such processes are not independent, from galaxy to galaxy, and quasi-random it may not match the random sampling.  However, our main use for these simulations is to produce covariance matrices and rough estimates of the mean halo mass of our samples. These depend primarily on the large-scale clustering and number density, and are much less sensitive to the details of our modeling.  Given the current uncertainties in our data, we believe these mocks are fit for this purpose. 

\begin{table}
\centering
\caption{Best-fit HOD parameters for each model and redshift bin.The AIC is defined as $= 2k + \chi^2_{\rm min}$ with $k$ the number of free parameters and lower numbers indicating preference for that model.}
\label{tab:bestfit}
\setlength{\tabcolsep}{8pt}
\begin{tabular}{l|ccc|ccc}
\hline\hline
 & \multicolumn{3}{c|}{$z=2.5$} & \multicolumn{3}{c}{$z=3.0$} \\
 & Standard & HMQ & Vel.\ bias & Standard & HMQ & Vel.\ bias \\
\hline
$\log M_{\mathrm{cut}}$ & $11.75$ & $11.75$ & $11.75$ & $11.75$ & $12.0$ & $11.0$ \\
$\log M_1$ & $12.45$ & $12.75$ & $12.45$ & $12.75$ & $13.0$ & $11.70$ \\
$\sigma$ & $0.66$ & $1.0$ & $0.66$ & $0.66$ & $1.0$ & $0.33$ \\
$\kappa$ & $0.33$ & $0.50$ & $1.0$ & $0.33$ & $0.50$ & $1.0$ \\
$\alpha$ & $0.33$ & $0.33$ & $1.0$ & $0.50$ & $1.0$ & $0.33$ \\
$Q$ & --- & $100.0$ & --- & --- & $100.0$ & --- \\
$\gamma$ & --- & $1.0$ & --- & --- & $1.0$ & --- \\
$p_{\mathrm{max}}$ & --- & $0.66$ & --- & --- & $0.66$ & --- \\
$\alpha_c$ & --- & --- & $1.0$ & --- & --- & $0.33$ \\
$\alpha_s$ & --- & --- & $0.10$ & --- & --- & $0.10$ \\
\hline
$f_\mathrm{sat}$ & $0.15$ & $0.22$ & $0.05$ & $0.07$ & $0.03$ & $0.24$ \\
$\bar{n}_\mathrm{HOD}/\bar{n}_\mathrm{LAE}$ & $90$ & $68$ & $81$ & $70$ & $26$ & $234$ \\
\hline
$\chi^2/\nu$ & $26.0/15$ & $25.8/12$ & $25.2/13$ & $23.9/15$ & $23.7/12$ & $23.8/13$ \\
 &  $=$\bf1.73 & $=2.15$ & $=1.94$ & $=$\bf1.59 & $=1.97$ & $=1.83$ \\
PTE & 0.038  & 0.011 & 0.022 & 0.067 & 0.022 & 0.033 \\
AIC & $36.0$ & $41.8$ & $39.2$ & $33.9$ & $39.7$ & $37.8$ \\
\hline\hline
\end{tabular}
\end{table}

\subsection{Standard model}
\label{sec:standard}

We start with the ``standard'' HOD functional form \cite{Zheng07,Wechsler18}.  Following standard practice, we assume all galaxies are hosted by halos and that the occupancy depends only upon the halo mass at the time of light emission.  We further assume that the distribution is specified entirely in terms of the mean occupancy. For each halo, the model assigns central galaxies with a binomial distribution and a Poisson-distributed number of satellites (chosen independently), with means
\begin{align}
\langle N_{\text{cen}}(M) \rangle &= \frac{1}{2}  \text{erfc} \left( \frac{\ln{M_{\text{cut}}/ M}}{\sqrt{2}\sigma} \right)
\label{eqn:hod-cen} \\
\langle N_{\text{sat}}(M) \rangle &= \langle N_{\text{cen}}(M) \rangle \left( \frac{M - \kappa M_{\text{cut}}}{M_1} \right)^{\alpha}
\quad \text{for } M > \kappa M_{\text{cut}}
\label{eqn:hod-sat}
\end{align}
where $M$ is the halo mass and \{$M_{\rm cut}$, $M_1$, $\sigma$, $\kappa$, $\alpha$\} are model parameters.
In this framework, $M_\mathrm{cut}$ defines the minimum halo mass required to host LAEs, while $\sigma$ controls the steepness of the threshold. $M_1$ marks the halo mass at which, on average, one satellite galaxy resides in a halo, while $\kappa$ suppresses satellite occupancy in lower-mass halos. 
We enumerate over a grid of 1680 models, over the parameter ranges $\log_{10} M_{\rm cut} \in \{11.00, 11.25, 11.50, 11.75, 12.00, 12.25, 12.50\}$, $\alpha \in \{0.33, 0.50, 0.66, 1.00, 1.33, 1.50\}$, $\sigma \in \{0.1, 0.33, 0.50, 0.66, 1.0\}$, $\kappa \in \{0.33, 0.66, 1.00\}$, and $M_1/M_{\rm cut} \in \{5, 10\}$, inspired by refs.~\cite{Yuan22,White24,Ebina25}. Of the models in this grid, we avoid models with $\log_{10}(\kappa M_\mathrm{cut})<11$ due to simulation resolution.

\subsection{HMQ model} \label{sec:HMQ}

We next consider a model with Halo Mass Quenching (HMQ) \cite{Alam20}, which has been shown to be mildly favored by the projected clustering of medium-band selected LAEs \cite{Ebina25}.  
In this model, the mean occupation of central galaxies is modified to include an additional quenching term that suppresses the occupation in halos above a characteristic mass scale. This is motivated by the property that large halos tend to have more neutral hydrogen in their circumgalactic medium, which can suppress the Ly$\alpha$ emission and thus the LAE selection \cite{Khoraminezhad25}. LAE studies involving hydrodynamical simulations have also shown simulation results favoring a similar, quenched distribution of central galaxies \cite{Sullivan25}. The mean occupation functions for satellites and centrals are
\begin{align}
    \langle N_{\text{cen}}(M) \rangle &= 2A\phi(M)\Phi(\gamma M) + \frac{1}{2Q}\left[ 1+ \text{erf}\left(\frac{\log_{10}{M/M_{\rm cut}}}{0.01}   \right)  \right]  \\
    \langle N_{\text{sat}}(M) \rangle &= \left( \frac{M - \kappa M_{\text{cut}}}{M_1} \right)^{\alpha}
    \quad \text{for } M > \kappa M_{\text{cut}}
\end{align}
where $M$ denotes the halo mass, and $\phi$ and $\Phi$ represent the normal probability density function and its corresponding cumulative distribution function, respectively
\begin{equation}
    \phi(x) = \mathcal{N}(\log_{10}{M_{\rm cut}},\sigma_M) \quad , \quad 
    \Phi(x) = \int_{-\infty}^x \phi(t)\,dt = \frac{1}{2}\left[1+\text{erf}\left(\frac{x}{\sqrt{2}}\right) \right]
    \quad .
\end{equation}
The constant $A = [p_{\rm max} - 1/Q] / \max[2\phi(x)\Phi(\gamma x)]$.  The model has four free parameters $\{\sigma_M, p_{\rm max}, \gamma, Q\}$, where $\sigma_M$ replaces the $\sigma$ parameter of the previous functional form. Among these, $p_{\rm max}$ and $\sigma_M$ govern the peak amplitude and width of the central occupation function, $\gamma$ determines the sharpness of the low-mass turn-on, and $Q$ sets the asymptotic value of $\langle N_{\rm cen} \rangle$ for $M \to \infty$. Here we enumerate over a grid of 3024 models, over the parameter ranges $\log_{10} M_{\rm cut} \in \{11.00, 11.25, 11.50, 11.75, 12.00, 12.25, 12.50\}$, $\alpha \in \{0.33, 0.66, 1.00\}$, $\kappa \in \{0.5, 1.0\}$, $\sigma_M \in \{0.33, 0.66, 1.0\}$, $M_1/M_{\rm cut} \in \{10, 30, 100\}$, $Q \in \{20, 100\}$, $\gamma \in \{1, 5\}$, and $p_{\rm max} \in \{0.33, 0.66\}$, inspired by refs.~\cite{Rocher23,Ebina25}. Similar to the standard model, we avoid models with $\log_{10}(\kappa M_\mathrm{cut})<11$ due to simulation resolution.

\subsection{Velocity bias}

Finally, we consider a model with velocity bias, which allows the velocity distribution of galaxies to differ from that of the halos or the dark matter particles in the host halos. 
We parametrize the velocity bias using two parameters $\alpha_c$ and $\alpha_s$ for centrals and satellites, respectively, such that \cite{Guo15}
\begin{align}
    \sigma_c &= \alpha_c \sigma_h \\
    v_s - v_h &= \alpha_s (v_p - v_h)
\end{align}
where $\sigma_c$ is the velocity dispersion of central galaxies, $\sigma_h$ is the velocity dispersion of the host halo, $v_s$ is the velocity of satellite galaxies, $v_h$ is the velocity of the host halo, and $v_p$ is the velocity of dark matter particles in the halo. We will enumerate over a grid of 2520 models, over the parameter ranges $\log_{10} M_{\rm cut} \in \{11.00, 11.25, 11.50, 11.75, 12.00, 12.25, 12.50\}$, $\alpha \in \{0.33, 0.66, 1.00\}$, $\sigma \in \{0.33, 0.66, 1.0\}$, $\kappa \in \{1.00\}$, $M_1/M_{\rm cut} \in \{5, 10\}$, $\alpha_c \in \{0, 0.33, 0.66, 1.0\}$, and $\alpha_s \in \{0.1, 0.5, 1.0, 1.5, 2.0\}$, following the grid in ref.~\cite{Guo15}. 

\subsection{Best-fit models and comparison}

The best-fit parameters for each model and redshift are shown in Table \ref{tab:bestfit}, along with derived statistics and goodness of fit. 
In both $z$-bins the best-fit $\chi^2$ is similar between HOD models, which leads to a noticeable preference for the standard model in both reduced $\chi^2$ and AIC due to the least number of parameters. 
This is reflected in the best-fit model correlation functions shown in Fig.~\ref{fig:HOD_fit}, which largely overlap within the fit range. The figure also demonstrates that the models can diverge in the one-halo regime ($s\lesssim 1\,h^{-1}\,\mathrm{Mpc}$), which is only partially included in this work.  These bins are plagued by small-number statistics given size of our dataset, making the estimation of covariance difficult. 

\begin{figure}
    \centering
    \includegraphics[width=\textwidth]{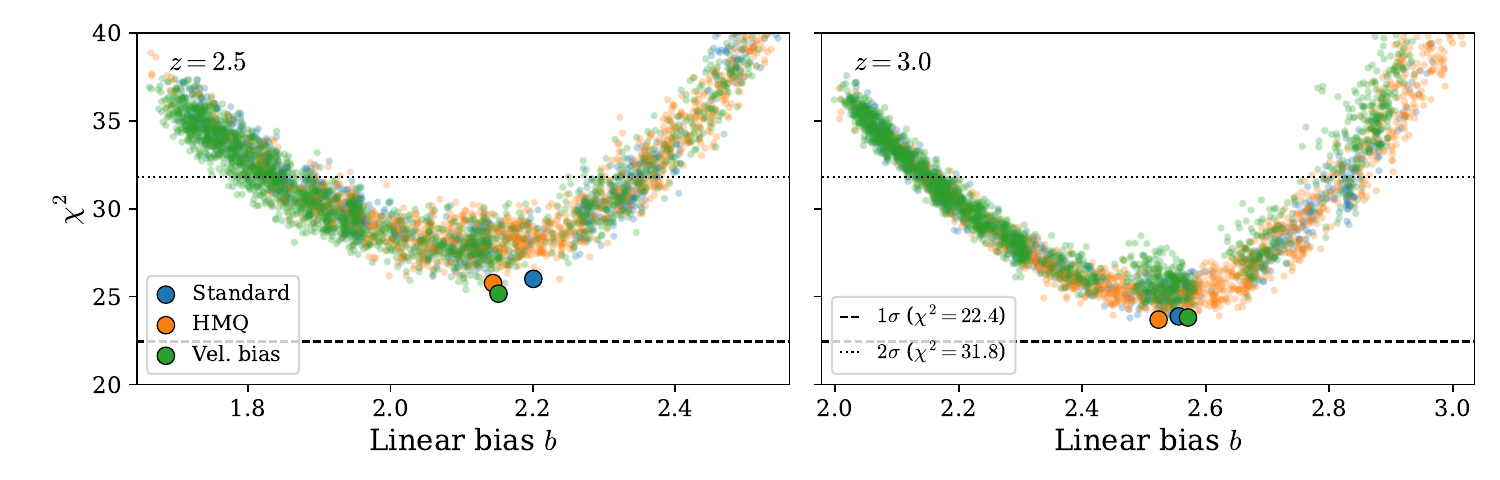}
    \caption{The distribution of $\chi^2$ values and linear bias $b$ for the different HOD models at $z=2.5$ and $z=3$. The relatively tight distribution indicates that the goodness of fit is strongly correlated with the bias. The termination of the distribution on the lower-bias end of $z=2.5$ bin reflects the mass resolution limit of the simulations used in this work.  The horizontal dashed and dotted lines show the $1$ and $2\,\sigma$ bounds for $\chi^2$ with $N_{\rm dof}=N_{\rm data}$.
    }
    \label{fig:chi2_bias}
\end{figure}

The influence of large-scale power on the fits is also demonstrated by the dependence of $\chi^2$ as a function of linear bias $b$, as shown in Fig.~\ref{fig:chi2_bias}. In both $z$-bins the $\chi^2$ follows a parabolic distribution, with best-fit models around $b\sim2.2$ and $\sim2.55$ for $z=2.5$ and $z=3$ bins, respectively.
Both biases are consistent within 1$\sigma$ with the linear biases derived from the best-fit model of ref.~\cite{Ebina25}, as well as with the Fourier-space fits using analytical theory in \S\ref{sec:analytical_fits} to $\lesssim0.5\sigma$. 
Note, however, that this constraint is rather non-trivial. The fits above use correlation function multipoles in a limited scale range, distinct from where one would typically obtain large-scale modes. Thus, we are fitting for smaller-scale features and extrapolating using the model, which at best mimics the underlying physics. 

The reduced $\chi^2$ values of $\sim1.6-1.7$ in both redshift bins reflect the difficulty of capturing the clustering with a simulation-based approach. As noted above, the HOD framework is phenomenological, mimicking the underlying physics of galaxy formation and imaging-based selection rather than reproducing it exactly, and these limitations may be degrading our fit. The Monte Carlo-based covariance may also underestimate the sample variance, for example through variation in the galaxy number density. Finally, the downsampling used to match the galaxy redshift distribution does not exactly reproduce the imaging-based selection applied to the data.

The mock-based approach nonetheless grants us several benefits. The most important for this work is the confirmation that fiber-assignment effects can be ignored, as detailed in Appendix \ref{app:fiberassign}. A second is agreement with the analytical theory fits (\S\ref{sec:analytical_fits}), which match our linear bias $b$ to $\lesssim 0.5\sigma$. Finally, the mock catalogs provide an alternative route to the Fourier-space covariance, which we use to validate the analytical covariances.

\begin{figure}
    \centering
    \includegraphics[width=0.98\linewidth]{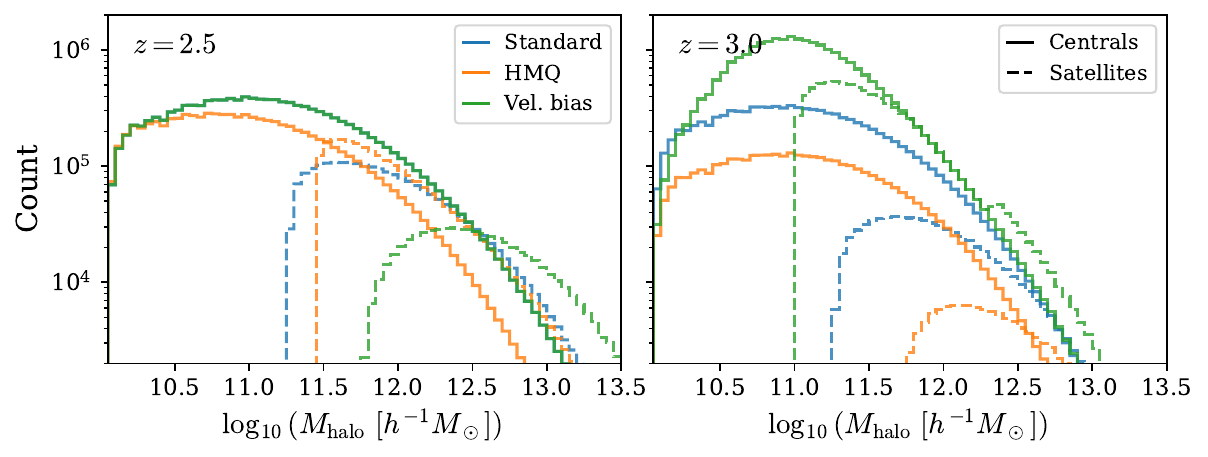}
    \caption{Halo occupation of HOD best-fit models shown in Table \ref{tab:bestfit}. The colors indicate the HOD model, and the line style indicate central and satellite galaxies.  The occupation of central galaxies of the standard model completely overlaps that of the velocity bias model at $z=2.5$, due to the shared $\log M_\mathrm{cut}$ and $\sigma$ between the bestfit models.}
    \label{fig:halo}
\end{figure}

Beyond the clustering itself, the HOD models also inform us about the halo occupancies of LAEs, as shown in Fig.~\ref{fig:halo}. At $z=2.5$, the central galaxies follow a similar distribution for all models, as evident by the consistent $\log M_\mathrm{cut}=11.75$, although the HMQ model exhibits fewer high-mass centrals due to the HMQ. The satellite distributions are roughly shared between the standard and HMQ models, where they predict moderate satellite occupancy near $\log_{10}M\sim 11.5$. The velocity bias model predicts a distribution that is less numerous and peaking at higher halo mass. This is reflected in the lower satellite fraction of the velocity bias best-fit model (Table \ref{tab:bestfit}). We caution that this best-fit model prefers rather extreme values of the velocity bias parameters, $\alpha_c$ and $\alpha_s$, though neither is particularly well constrained by our data.  Since $\alpha_s\sim 0.1$, the satellites behave in a similar way to centrals as far as small-scale velocity dispersion is concerned and this -- combined with our limited data -- opens a degeneracy direction in the modeling which is poorly constrained.
At $z=3$, each model predicts a different distribution of centrals and satellites. The central distribution of standard and HMQ models appear similar, although with a factor of $\sim2$ difference in number. That for the velocity bias model peaks sharply near $\log M\sim 11$, with the peak distribution more numerous than the standard distribution by a factor of $\sim3$. The velocity bias model also exhibits more satellites, which show similar distributions to the centrals above $\log M \sim 11.5$. The standard and HMQ models predict much fewer satellites, with higher minimum host halo mass. This is reflected in the high satellite fraction of the velocity bias model at $z=3$, in contrast to standard and HMQ models (Table \ref{tab:bestfit}).  The velocity bias model is able to accommodate a high satellite fraction without enhanced velocity dispersion by decreasing $\alpha_s$ to $\sim 0.1$.  This value is unlikely on physical grounds.

We must note that the satellite fraction constraints from this analysis is relatively weak, as satellite galaxies primarily contribute to small-scale (one-halo) correlations at $s\lesssim1\,h^{-1}\,\mathrm{Mpc}$, where we have only a few, loosely constraining data points. Nonetheless, it is modestly encouraging that the satellite fraction of most models are low, which typically correspond with a weak Fingers-of-God (FoG) effect. Further discussion on FoG and small-scale dynamics will be held in \S\ref{sec:FoG}.

\section{Fourier-space fits}
\label{sec:analytical_fits}

Here we perform analysis in Fourier space using the effective field theory (EFT) formalism that perturbatively models the quasi-linear modes of the power spectrum \cite{Bernardeau02,Ivanov22b}. 
While this does not provide information on the halo occupation as from the HOD fits in \S\ref{sec:HOD}, it provides a clear separation of model-valid scales and hence a more robust estimate of the galaxy bias, an important input for cosmology forecasts. Furthermore, it allows us to incorporate simplistic models of radiative transfer to estimate its impact. 

\subsection{Power spectrum theory}
\label{sec:ps_theory}

The EFT-based modeling of the power spectrum is the standard formalism used to perform cosmology fits in full-shape cosmology analysis \cite{Ivanov22b,DESI24-V}. The formalism describes the power spectrum as a combination of galaxy biases, counterterms (small-scale physics), and stochasticity
\begin{equation}
    P(k,\mu) = \sum_{X,Y} b_X b_Y P_{XY}(k,\mu) + P_{c.t.}(k,\mu) + P_{\rm stoch}(k,\mu)
\end{equation}
where $\mu$ is the cosine of the angle with the line-of-sight.
The first term is sourced from the galaxy bias expansion, which expands the galaxy density field $\delta_g$ using possible contractions of the underlying matter field $\delta$ (enumerating over $X$ and $Y$). To one-loop order, this is
\begin{equation}
    \delta_g = b_1 \delta + \frac{b_2}{2}\delta^2 + b_s s^2
    \label{eq:bias_expansion}
\end{equation}
where $s^2 = s_{ij}^2$ is the self-contraction of the shear field \cite{Desjacques18}. Here we have neglected the third order bias $b_3$ following standard practice \cite{DESI24-V}.  
The second term is the counterterms, which are sourced from small-scale interactions not captured in the (large-scale) equations of motion used to construct the perturbative terms. Including all terms allowed by the symmetry of the system, this becomes 
\begin{equation}
    P_{c.t}\supset\alpha_{2n}\, \mu^{2n}(k/k_\star)^{2}P_{L} \quad.
\end{equation}
The final term of the power spectrum arise from the stochasticity of the density field
\begin{equation}
    P_{s.t.} = N + \sum_{n=1} N_{2n}(k\mu)^{2n}
\end{equation}
where the leading term $N$ is shot noise and the following terms describe the Fingers-of-God (FoG) effect. We will analyze the FoG effect in detail in \S\ref{sec:FoG}.

It is noteworthy that in addition to symmetry-allowed small-scale dynamics, these counterterms and stochastic terms also absorb modest amounts of redshift error \cite{Chen20a,Maus24b}.  Redshift errors act to damp the non-stochastic contributions to the power spectrum by the squared modulus of the characteristic function of the error PDF.  For example for an exponential redshift-error PDF the clustering power is multiplied by $|\mathcal{D}(k_\parallel)|^2=[1+(k_\parallel\sigma_z)^2/2]^{-2}$ with $k_\parallel = k\mu$.  For a Gaussian $|\mathcal{D}(k_\parallel)|^2=\exp[-(k_\parallel\sigma_z)^2]$. 
In either case, and in general, the Taylor expansion of $k\mu\sigma_z$ can be captured by a redefinition of the counterterms and FoG parameters, up to orders consistent with the perturbative formalism, allowing us to not include any further degrees of freedom.  As long as the impact of redshift errors are small, we should not need to recenter or modify our priors on the nuisance parameters. We further confirm in \S\ref{sec:FoG} that any $\mu$-dependent damping is small for LAEs, confirming this choice. 

The above formalism describes the theoretical setup for the galaxy auto-spectrum. Here we also consider the cross-correlation of LAE and DESI DR2 QSOs as well, which requires a different treatment. 
The bias-based term is a natural extension, where one considers all combinations of the bias expansion (Eqn.~\ref{eq:bias_expansion}), hence introducing no new parameters. The counterterms can also be predicted from already existing auto-spectrum parameters
\begin{equation}
    \alpha_{2n}^\times = \frac{1}{2}\sum_{m=0}^{n}\left[\alpha_{2m}^a \, c_{n-m}(b_1^a, b_1^b) + \alpha_{2m}^b \, c_{n-m}(b_1^b, b_1^a)\right],
\end{equation}
where $a$ and $b$ label the tracer, 
\begin{equation}
    c_0(b_1^X, b_1^Y) = \frac{b_1^Y}{b_1^X}\,, \qquad c_k(b_1^X, b_1^Y) = (-1)^{k+1}\!\left(\frac{f}{b_1^X}\right)^{\!k}\!\left(1 - \frac{b_1^Y}{b_1^X}\right) \;\text{for } k\geq 1\,,
\end{equation}
and $b_1^X$ denotes the linear bias of tracer $X$ \cite{Mergulhao23,Ebina24}. The first two counterterms then become
\begin{align}
    \alpha_0^\times &= \frac{1}{2}\!\left(\alpha_0^a\frac{b_1^b}{b_1^a} + \alpha_0^b\frac{b_1^a}{b_1^b}\right), \\
    \alpha_2^\times &= \frac{1}{2}\!\left(\alpha_2^a\frac{b_1^b}{b_1^a} + \alpha_2^b\frac{b_1^a}{b_1^b} + \frac{f}{b_1^a}\alpha_0^a\!\left(1-\frac{b_1^b}{b_1^a}\right) + \frac{f}{b_1^b}\alpha_0^b\!\left(1-\frac{b_1^a}{b_1^b}\right)\right)
\end{align}
where, in the context of this work, $a=\mathrm{LAE}$ and $b=\mathrm{QSO}$. The stochastic term, on the other hand, gives rise to a new set of parameters that contribute
\begin{equation}
    P_{s.t.}^\times = N^\times + \sum_{n=1} N_{2n}^\times(k\mu)^{2n}
\end{equation}
as the cross-stochasticity of tracers cannot be predicted from other parameters. We refer the reader to refs.~\cite{Mergulhao23,Ebina24} for details.

Using the above formalism through the \texttt{velocileptors} package \cite{Chen20a,Chen20c}, we fit the monopole and quadrupole of the LAE auto-spectrum and LAE-QSO cross-spectrum, as well as the QSO auto-spectrum monopole. 
The final set of parameters then becomes
\begin{align}
    \{&b_1^{\rm LAE}, \, b_2^{\rm LAE}, \, b_s^{\rm LAE}, \, \alpha_0^{\rm LAE}, \, \alpha_2^{\rm LAE}, \, N^{\rm LAE}, \, N_2^{\rm LAE} \nonumber\\
    &b_1^{\rm QSO}, \, b_2^{\rm QSO}, \, b_s^{\rm QSO}, \, \alpha_0^{\rm QSO}, \, \alpha_2^{\rm QSO}, \, N^{\rm QSO}, \, N_2^{\rm QSO},N^\times, \, N_2^\times\},
\end{align}
where we neglect the small-scale parameters that only arise at the hexadecapole level and hence are degenerate for the monopole and quadrupole. With cosmology fixed to that from \textit{Planck} \cite{PCP18}, we perform the fits to $k_\mathrm{max}=0.3\,h\,\mathrm{Mpc}^{-1}$, which is sufficiently below the non-linear scale $k_{\rm nl}\sim0.5\,h\,\mathrm{Mpc}^{-1}$ at $z=3$ \cite{Chen20a,Sailer21}. We have further validated that the fits are consistent with a lower scale cut ($k_\mathrm{max}=0.25\,h\,\mathrm{Mpc}^{-1}$), as well as with a fit using simpler, linear-theory \cite{Kai87} + phenomenological FoG model \cite{Jackson72,Peacock92,Peacock94,Park94,Ballinger96,Ham98,Hatton99,White01,Scoccimarro04} with $k_\mathrm{max}=0.2\,h\,\mathrm{Mpc}^{-1}$. 

The theory multipoles $\ell \in \{0, 2, 4\}$ are convolved with the
survey window matrix (\S\ref{sec:measurement}) before comparison to the data:
\begin{equation}
    P_\ell^{\rm obs}(k) = \sum_{\ell'} \int \mathrm{d}k'\, W_{\ell\ell'}(k, k')\, P_{\ell'}^{\rm th}(k') \,.
\end{equation}
The linear matter power spectrum $P_{mm}(k)$ is computed at the effective redshift of each spectrum.

The parameters are fit by a Markov-Chain Monte Carlo (MCMC), using the \texttt{cobaya} package and a $R-1=0.01$ ``Gelman-Rubin'' convergence condition. The likelihood is constructed using the analytical covariance matrix discussed in \S\ref{sec:measurement}. We employ priors on LAE parameters informed from simulation fits to LAE clustering data in \S\ref{sec:HOD}, as well as those in refs.~\cite{Ebina25,Sullivan25}. For the QSO parameters, we adopt priors based on fits to DESI QSO CoLoRe-2LPT mocks \cite{RuizHerrera26} to compensate for the lack of QSO quadrupole, although the fits to large-scale parameters turn out to be consistent with wide uninformative priors.
With both tracers, we use wide uniform priors for linear bias. The full set of priors are shown in Table~\ref{tab:eft_priors}. 

\begin{table}
\centering
\caption{Priors on EFT parameters. $\mathcal{N}(\mu, \sigma)$ denotes a Gaussian prior; $\mathcal{U}(a, b)$ a uniform prior. QSO second order biases ($b_2$, $b_s$) and counterterms ($\alpha_0$, $\alpha_2$) are informed by DESI CoLoRe-2LPT mock posteriors.}
\renewcommand{\arraystretch}{1.2}
\begin{minipage}[t]{0.35\textwidth}
\vspace{0pt}
\centering
\begin{tabular}{l|c}
\hline
Parameter & Prior \\
\hline
$b_{1}^\mathrm{LAE}$ & $\mathcal{U}(0.5,\, 6.0)$ \\
$b_{2}^\mathrm{LAE}$ & $\mathcal{N}(0,\, 1)$ \\
$b_{s}^\mathrm{LAE}$ & $\mathcal{N}(0,\, 2)$ \\
$\alpha_{0}^\mathrm{LAE}$ & $\mathcal{N}(-5,\, 5)$ \\
$\alpha_{2}^\mathrm{LAE}$ & $\mathcal{N}(0,\, 7)$ \\
$N^\mathrm{LAE}$ & $\mathcal{N}(\bar{n}_\mathrm{LAE}^{-1},\, \bar{n}_\mathrm{LAE}^{-1})$ \\
$N_{2}^\mathrm{LAE}$ & $\mathcal{N}(0,\, 5\times10^5)$ \\
$\alpha$ & $\mathcal{U}(-5,\, 5)$ \\
& \\
\hline
\end{tabular}
\end{minipage}
\hfill
\begin{minipage}[t]{0.6\textwidth}
\vspace{0pt}
\centering
\begin{tabular}{l|cc}
\hline
Parameter & Prior ($z=2.5$) & Prior ($z=3.0$) \\
\hline
$b_{1}^\mathrm{QSO}$ & \multicolumn{2}{c}{$\mathcal{U}(1.5,\, 6.0)$} \\
$b_{2}^\mathrm{QSO}$ & $\mathcal{N}(1.4,\, 2.8)$ & $\mathcal{N}(4.3,\, 5.5)$ \\
$b_{s}^\mathrm{QSO}$ & $\mathcal{N}(7.2,\, 3.0)$ & $\mathcal{N}(-3.9,\, 6.7)$ \\
$\alpha_{0}^\mathrm{QSO}$ & $\mathcal{N}(-5.0,\, 34.5)$ & $\mathcal{N}(15.3,\, 38.4)$ \\
$\alpha_{2}^\mathrm{QSO}$ & $\mathcal{N}(39.5,\, 29.7)$ & $\mathcal{N}(21.1,\, 36.9)$ \\
$N^\mathrm{QSO}$ & \multicolumn{2}{c}{$\mathcal{N}(\bar{n}_\mathrm{QSO}^{-1},\, \bar{n}_\mathrm{QSO}^{-1})$} \\
$N_{2}^\mathrm{QSO}$ & \multicolumn{2}{c}{$\mathcal{N}(0,\, 5\times10^5)$} \\
$N^{\times}$ & \multicolumn{2}{c}{$\mathcal{N}(0,\, 1/\sqrt{\bar{n}_\mathrm{LAE}\,\bar{n}_\mathrm{QSO}})$} \\
$N_{2}^{\times}$ & \multicolumn{2}{c}{$\mathcal{N}(0,\, 5\times10^5)$} \\
\hline
\end{tabular}
\end{minipage}
\label{tab:eft_priors}
\end{table}

\begin{figure}
    \centering
    \includegraphics[width=\textwidth]{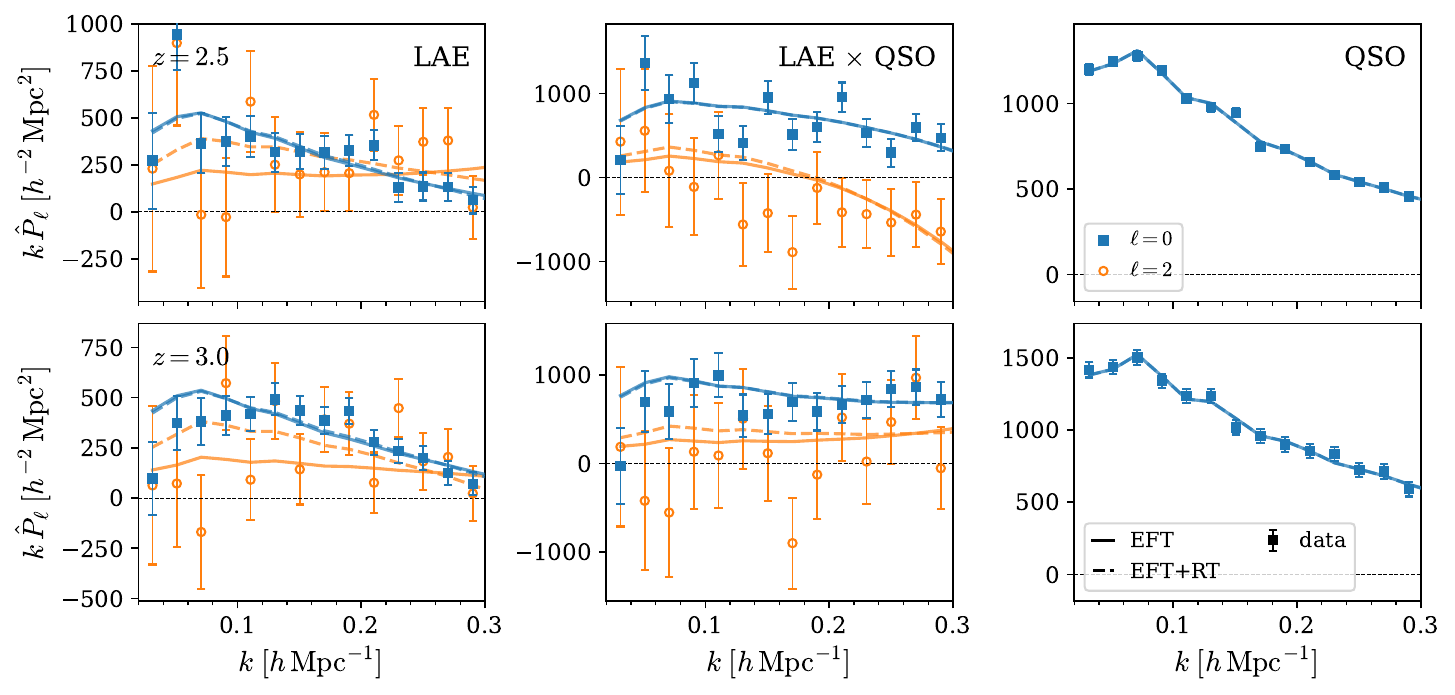}
    \caption{Best-fit EFT theory model (solid lines) compared to the measured power spectrum multipoles (markers with error bars) for the LAE auto-spectrum, LAE\,$\times$\,QSO cross-spectrum, and QSO auto-spectrum, with the Poisson shot noise subtracted from the auto-spectra ($\hat{P}_{ab}=P_{ab}-\delta^K_{ab}/\bar{n}_a$ for tracers $a$, $b$). The dashed lines show the best-fit model with the RT contribution ($\alpha$). Each row corresponds to a different redshift bin. The best-fit parameters are listed in Table~\ref{tab:eft}.  
    }
    \label{fig:pk_fit}
\end{figure}

\begin{figure}
    \centering
    \includegraphics[width=0.98\linewidth]{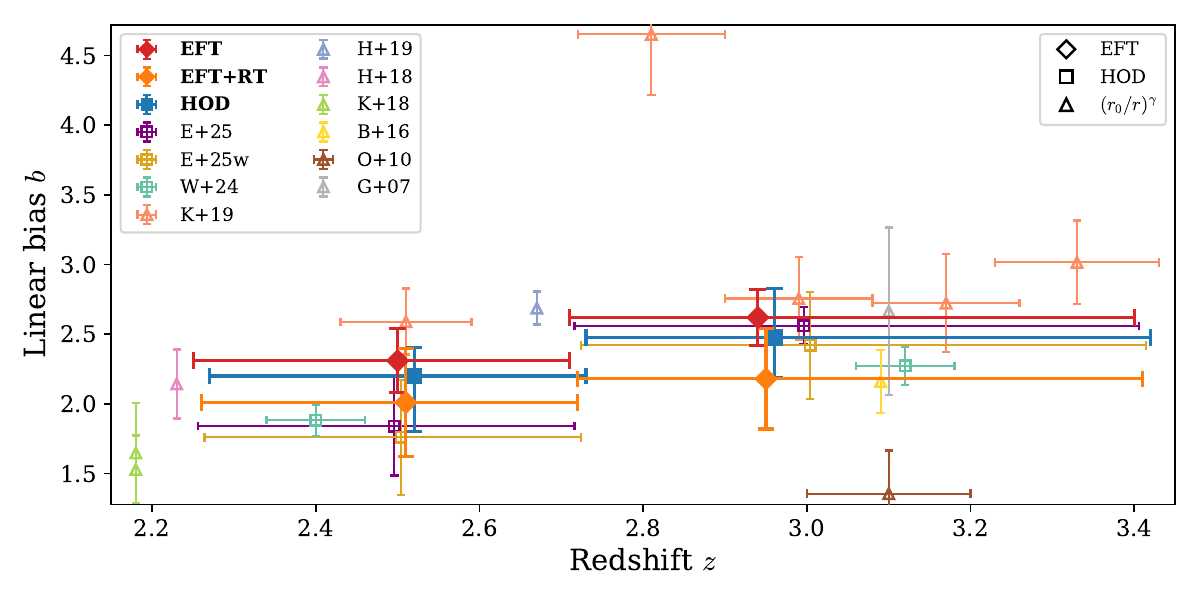}
    \caption{The linear bias estimates in this work against past work \cite{Gawiser07,Ouchi10,Bielby16,Kusakabe18, Hao18, Hong19,Khostovan19,White24,Ebina25}, each labeled in the legend; the E+25w refer to the ``wide'' sample of ref.~\cite{Ebina25}. The errors on the HOD fits (\S\ref{sec:HOD}) are derived using the $\Delta\chi^2$ from the best-fit model, within the standard HOD model.}
    \label{fig:b_vs_z}
\end{figure}

The fit results are shown in Fig.~\ref{fig:pk_fit} and Table \ref{tab:eft}. The LAE linear bias at each redshift is $b_1^\mathrm{LAE}= 2.31 \pm 0.23$ and $2.62 \pm 0.20$ at $z=2.5$ and 3.0, respectively, providing the best constraints on the linear bias of medium-band selected LAEs to date.
These values are consistent with the values derived using HOD models in \S\ref{sec:HOD} to $\lesssim0.5\sigma$, as well as previous fits in the literature aside from a limited set of outliers (Fig.~\ref{fig:b_vs_z}). Due to the difference in LAE selection and analysis methods, the latter agreement is not expected a priori, as discussed in \S\ref{sec:tracers}. Even the former is not expected, due to the correlation function range in the HOD analysis sampling a different set of modes compared to those used here. Nonetheless, the agreement attained is encouraging. 

\begin{figure}
    \centering
    \includegraphics[width=0.97\linewidth]{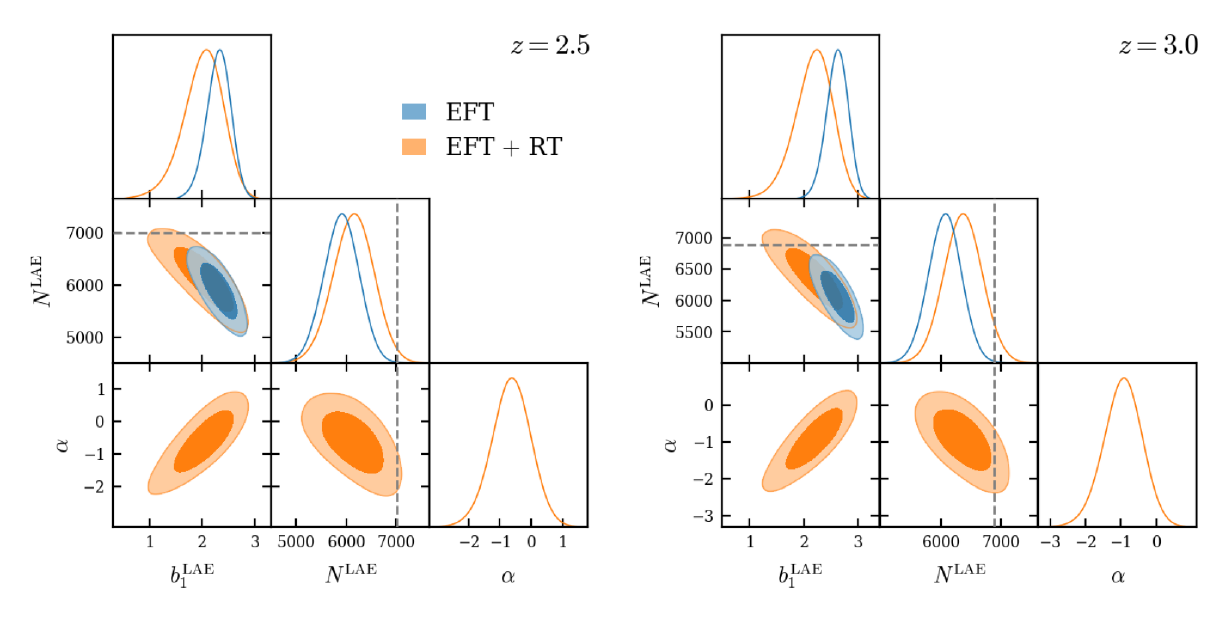}
    \caption{The posterior distribution of LAE linear bias $b_1$, shot noise $N$, and radiative transfer $\alpha$ (if applicable) in the EFT fits. The dashed lines demonstrate the Poisson shot noise prediction.}
    \label{fig:contour}
\end{figure}

The posterior contours shown in Fig.~\ref{fig:contour} provide further insight into the fits. We note the degeneracy between $b_1^\mathrm{LAE}$ and $N^\mathrm{LAE}$. This is simply due to the most stringent constraint on LAE clustering coming from the LAE auto-spectrum monopole on scales where shot noise is not negligible, which fixes a combination of bias and shot noise. Another is the small ($\sim 1/7\bar{n}$) deviation of $N^\mathrm{LAE}$ from the Poisson shot noise value.  This suggests that deviations from Poisson shot noise are small, but likely non-zero.

The EFT fits also constrain the ratio of clustering power to stochastic power for these LAE samples. Using linear theory we find $\bar{n}P_0(k=0.1\,h\,\mathrm{Mpc}^{-1})=0.87$ and 0.85 for each redshift bin, close to but slightly smaller than the ``optimal'' value of $\bar{n}P=1$ for these scales.

\begin{table}[h]
\centering
\caption{EFT model fits for the joint LAE auto $P_0+P_2$ + LAE$\times$QSO cross $P_0+P_2$ + QSO auto $P_0$ ($k_{\rm max}=0.30\,h\,{\rm Mpc}^{-1}$). The $+$RT rows additionally include $\alpha = 1 - f_{\rm LAE}/f_{\rm fid}$, where $\alpha = 0$ in the absence of RT or complex selection effects. The AIC is defined as $= 2k + \chi^2_{\rm min}$ with $k$ the number of free parameters and lower numbers indicating preference for that model.}
\begin{tabular}{llccccc}
\hline\hline
$z$-range & Fit & $b_{\rm LAE}$ & $b_{\rm QSO}$ & $\alpha$ & $\chi^2/\nu$ & AIC \\
\hline
2.26--2.72 & EFT & $2.31 \pm 0.23$ & $3.43 \pm 0.12$ & \multicolumn{1}{c}{$-$} & $75.5/54 = 1.40$ & $107.5$ \\
 & $+$RT & $2.01 \pm 0.39$ & $3.42 \pm 0.13$ & $-0.64 \pm 0.63$ & $75.0/53 = 1.41$ & $109.0$ \\
\hline
2.72--3.41 & EFT & $2.62 \pm 0.20$ & $4.44 \pm 0.13$ & \multicolumn{1}{c}{$-$} & $61.4/54 = 1.14$ & $93.4$ \\
 & $+$RT & $2.18 \pm 0.36$ & $4.44 \pm 0.13$ & $-0.95 \pm 0.55$ & $59.8/53 = 1.13$ & $93.8$ \\
\hline\hline
\end{tabular}
\label{tab:eft}
\end{table}

\subsection{Radiative Transfer}
\label{sec:RT}

An important complication specific to LAEs is the possible impact of radiative transfer (RT) of Ly$\alpha$ photons through the neutral hydrogen in the circumgalactic and intergalactic medium.
While there is not a simple analytic model for the RT contribution to the clustering, given the quality of our data we choose to parametrize it as an additional $\mu$-dependent contribution that mimics RSD at leading order \cite{Zheng11, Ebina24}. We model this using the parameter $\alpha$, so that the LAE linear density field becomes 
\begin{equation}
    \delta_{\rm LAE} = \left[ b+(1-\alpha)f\mu^2\right] \delta_L + \epsilon
\end{equation}
where $\delta_L$ is the linear density field and $\epsilon$ is the stochastic field. This therefore sets $f_{\rm LAE}=(1-\alpha)f$, making the RSD signal tracer dependent.
Because $\alpha$ enters multiplying $f$, in a cosmology fit it is fully degenerate with the growth rate from the linear LAE auto-spectrum alone; breaking this degeneracy requires either higher-order information, external information on $f$ or a cross-correlation with a tracer unaffected by RT (e.g.\ QSOs, LBGs, or CMB lensing).

We perform fixed cosmology fits here, but nonetheless will employ cross-correlation with DESI DR2 QSOs \cite{DESI-DR2} to tighten our measurement of $\alpha$. The fit results are shown in Figs.~\ref{fig:pk_fit} and \ref{fig:contour}, and Table \ref{tab:eft}. The results indicate a $<2\sigma$ deviation from $\alpha=0$ in both $z$-bins, placing constraints on the strength of RT at order $\sigma(\alpha)\sim 0.5$. 
Of the two $z$-bins, the deviation from $\alpha=0$ at $z=3$ is $1.7\sigma$. This is an interesting result, but is still consistent with the simplest case that RT is negligible and this is reflected in the AIC preferring neither model strongly at either $z$-bin. If the RT is confirmed to be non-zero it will shine light into the complex scattering of Ly$\alpha$ photons and have significant implications for future high-$z$ cosmology campaigns. It will be important to follow this up in the future with a larger dataset, with more constraining power. 

Besides the one-dimensional posterior on $\alpha$, the contours in Fig.~\ref{fig:contour} provide some insight into the parameter relations in the fit. We easily find that there is a strong correlation between $\alpha$ and $b_1$, which is expected from the amplitude degeneracy in the auto-spectrum. This is the (weakened) degeneracy after including a second tracer (QSOs), highlighting the importance of cross-correlation when constraining RT. 
By extension of the $\alpha$-$b_1$ degeneracy, we find that $\alpha$ is also weakly degenerate with $N$ due to the tight $P_0$ constraint inducing a $b_1$-$N$ degeneracy, as discussed in \S\ref{sec:ps_theory}.

\section{Fingers-of-God and non-linear effects} \label{sec:FoG}

Fingers-of-God (FoG) effects arise from the peculiar velocities of galaxies within virialized structures, elongating clustering signals along the line of sight in redshift space and suppressing the observed power spectrum on small scales \cite{Jackson72}. 
As a direct consequence of these effects arising from small-scale physics, FoG are stochastic processes that decorrelate the galaxy clustering from the primordial matter field, restricting the cosmological output from galaxy surveys at small scales. 
This also counters part of the motivation for conducting cosmology at high-$z$, which is that the clustering signal is more linear and thus more easily modeled to smaller scales. The stronger biases of high-$z$ galaxies may further strengthen FoG effects as well. 
For next-generation surveys such as DESI-II, accurately characterizing and mitigating FoG effects is essential, as uncontrolled FoG systematics directly impact the constraining power on cosmological parameters.
In this section, we measure the impact of FoG effects on LAE clustering using several complementary approaches.

\subsection{Anisotropic correlation function}

\begin{figure}[htbp]
    \centering
    \begin{minipage}{0.5\textwidth}
        \centering
        \includegraphics[width=\linewidth]{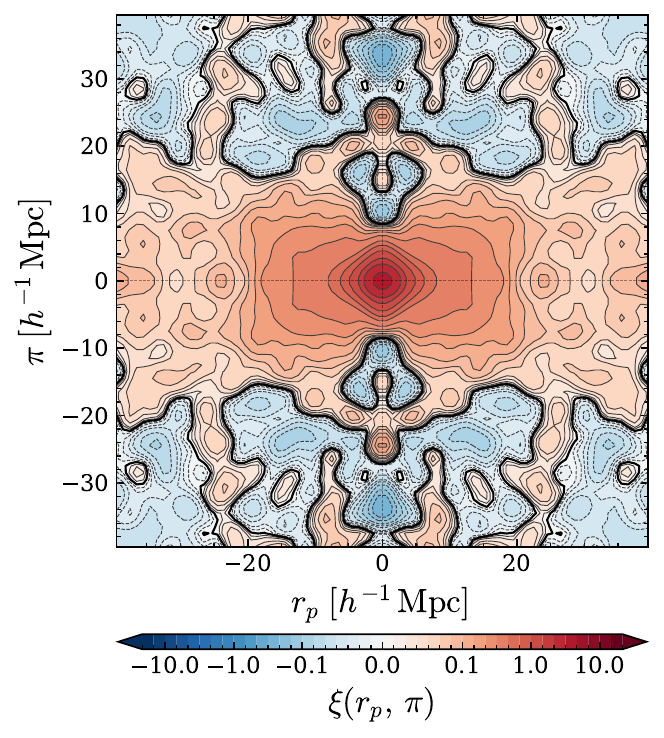}
    \end{minipage}
    \hfill
    \begin{minipage}{0.48\textwidth}
        \centering
        \includegraphics[width=\linewidth]{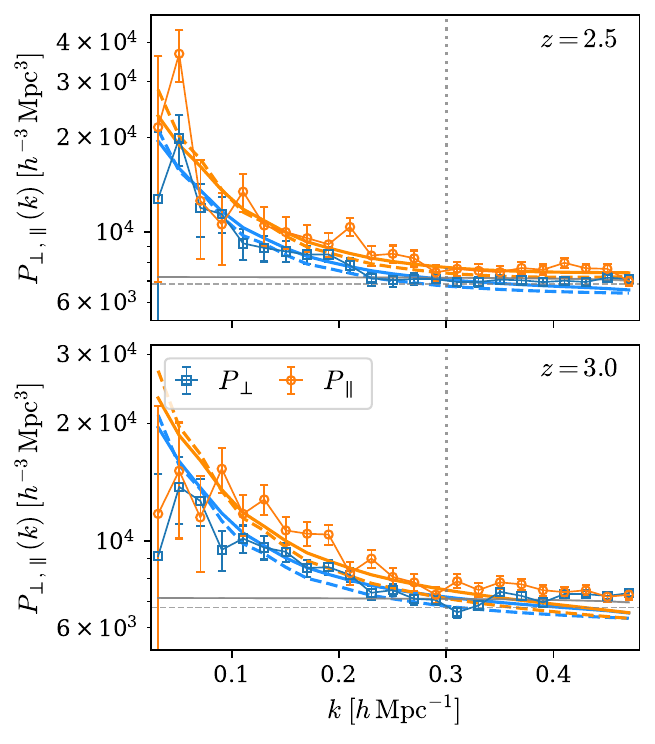}
    \end{minipage}
    \caption{\textbf{Left:} The anisotropic correlation function $\xi(r_p,\pi)$ for the combined LAE sample. The contours indicate log-spaced bins of $\xi(r_p,\pi)$. The elongation along the $\pi$ direction at small scales would be signature of FoG effects, while the compression at large scales is consistent with linear RSD.
    \textbf{Right:} The transverse ($P_\perp$) and line-of-sight ($P_\parallel$) power spectra for the LAE sample, defined by integrating $P(k,\mu)$ over $\mu < 2/3$ and $\mu > 2/3$, respectively. The (lack of) suppression of $P_\parallel$ at small scales constrains the strength of FoG effects. 
    This is plotted against the EFT theory fits derived in \S\ref{sec:analytical_fits}, with (solid) and without (dashed) window convolution. 
    The errors shown are analytical errors derived from the Gaussian covariance matrix used in EFT fits and hence should only used to guide the eye at small scales.
    The Poisson shot noise, with (solid) and without (dashed) window convolution, is shown with a horizontal gray line to demonstrate the strength of non-Poisson stochasticity. The dotted vertical line shows the $k_\mathrm{max}=0.3\,h\,\mathrm{Mpc}^{-1}$ of the EFT theory fits. 
    }
    \label{fig:anisotropy}
\end{figure}

The anisotropy of the correlation function $\xi(r_p,\pi)$, where $r_p$ and $\pi$ are the transverse and line-of-sight separations, respectively, provides a direct probe of FoG effects \cite{Fisher94}. On small scales, the correlation function is expected to be elongated along the $\pi$ direction due to the random motions of galaxies within halos (FoG), while at large-scales, the correlation function is compressed along the $\pi$ direction due to the RSD in linear theory \cite{Kai87}. The two regimes are visually apparent on the left panel of Fig.~\ref{fig:anisotropy}. We further quantify this using the power spectrum in \S\ref{sec:FoG_fit}.

\subsection{High multipoles in the power spectrum}
\label{sec:mucut}

\begin{figure}
    \centering
    \includegraphics[width=0.98\linewidth]{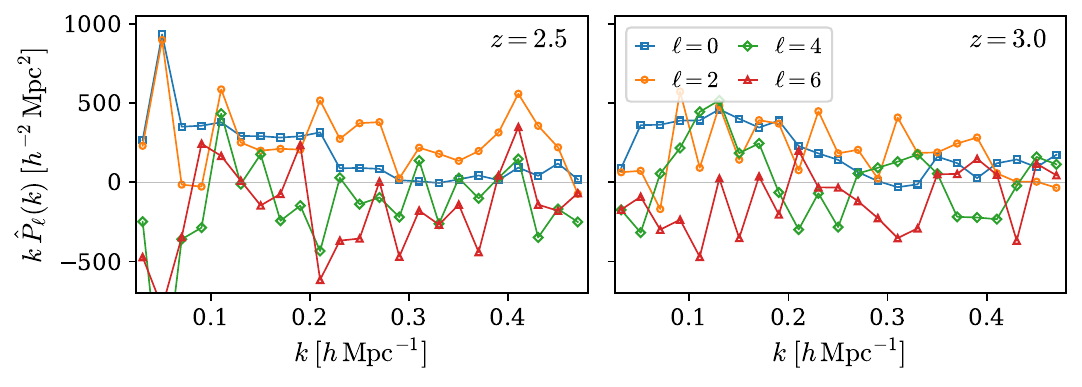}
    \caption{Multipoles of the LAE power spectrum with the shot noise subtracted ($\hat{P}$). This is the input to the wedge power spectra shown in the right panel of Fig.~\ref{fig:anisotropy}.  The errors on the higher multipoles are both large and highly correlated and have been suppressed for visual clarity. The high-$k$ of the zero-crossing, at $k\gtrsim0.45\,h\,\mathrm{Mpc}^{-1}$, indicates that the FoG suppression is weak \cite{BaleatoLizancos25}.}
    \label{fig:multipoles}
\end{figure}

As a complementary diagnostic, we construct ``power spectrum wedges'' by integrating $P(k,\mu)$ over restricted ranges of $\mu$ \cite{Kazin12}. 
For a given cut value $\mu_c$ (we adopt $\mu_c=2/3$), we define a transverse component $P_{\perp}(k)$ and a radial component $P_{\parallel}(k)$ 
\begin{equation}
    P_{\perp}(k) = \frac{1}{\mu_c}\int_0^{\mu_c} P(k,\mu)\,\mathrm{d}\mu
             = \sum_\ell a_\ell^{\perp}\, P_\ell(k) \,, \qquad
    P_{\parallel}(k) = \frac{1}{1-\mu_c}\int_{\mu_c}^{1} P(k,\mu)\,\mathrm{d}\mu
             = \sum_\ell a_\ell^{\parallel}\, P_\ell(k) \,,
\end{equation}
where
\begin{equation}
    a_\ell^{\perp} = \frac{1}{\mu_c}\int_0^{\mu_c} \mathcal{L}_\ell(\mu)\,\mathrm{d}\mu \,, \qquad
    a_\ell^{\parallel} = \frac{1}{1-\mu_c}\int_{\mu_c}^{1} \mathcal{L}_\ell(\mu)\,\mathrm{d}\mu \,,
\end{equation}
are coefficients obtained by integrating the Legendre polynomials,  $\mathcal{L}_\ell$, over $[0,\mu_c]$ and $[\mu_c,1]$, respectively.  For $\mu_c=2/3$, $a^{\perp}_\ell\simeq [1,-0.28,0,0.06]$ and $a^{\parallel}_\ell\simeq [1,0.56,0.02,-0.13]$. Because $P_{\parallel}$  preferentially weights modes along the line of sight, it is more sensitive to FoG suppression than either the monopole or the quadrupole alone, while $P_{\perp}$ is dominated by transverse modes and serves as a control that is largely insensitive to velocity dispersion effects. The input multipoles $P_\ell$ are shown in Fig.~\ref{fig:multipoles}.  Note that because the errors on the higher-order multipoles are so correlated, the sums $P_{\perp}$ and $P_{\parallel}$ are much smoother than the constituent multipoles.

Errors on the wedge spectra are propagated from the multipole covariance matrix,
\begin{equation}
    \sigma^2_{P_{\perp,\parallel}}(k)
      = \mathbf{a}^T_{\perp,\parallel}\, \mathbf{C}(k)\, \mathbf{a}_{\perp,\parallel} \,,
\end{equation}
where $\mathbf{C}(k)$ is the Gaussian-approximation linear-theory covariance used in \S\ref{sec:analytical_fits}.  This predicts that the errors on the two statistics are largely uncorrelated ($r\lesssim 0.1$).  This is expected.  For scales that are only modestly affected by the window function we can estimate the covariance of any two power spectrum multipoles, in a thin shell containing $N_{\rm mode}$ modes, as
\begin{equation}
    \mathrm{Cov}\left[ P_{\ell_1}(k) , P_{\ell_2}(k) \right]
    = \frac{1}{N_{\rm mode}}\sum_{\ell_3\ell_4} \kappa_{\ell_1\ell_2}^{\ell_3\ell_4} P_{\ell_3}P_{\ell_4}
    \quad , \quad
    N_{\rm mode} = \frac{4\pi\, k^2\,\Delta k}{(2\pi/L_{\rm box})^3}
\end{equation}
with
\begin{equation}
  \kappa_{\ell_1\ell_2}^{\ell_3\ell_4} \equiv (2\ell_1+1)(2\ell_2+1)
    \int_{-1}^{+1}d\mu\ \mathcal{L}_{\ell_1}(\mu)\mathcal{L}_{\ell_2}(\mu)
    \mathcal{L}_{\ell_3}(\mu)\mathcal{L}_{\ell_4}(\mu)
\end{equation}
so the wedge covariance follows by direct propagation,
\begin{equation}
    C_{ij}(k) = \frac{1}{N_{\rm mode}}
    \sum_{\ell_1,\ell_2} a_{\ell_1}^{(i)} a_{\ell_2}^{(j)}
    \sum_{\ell_3,\ell_4}
    \kappa_{\ell_1\ell_2}^{\ell_3\ell_4}\, P_{\ell_3} P_{\ell_4}\,,
    \qquad i,j\in\{\perp,\parallel\}\,,
    \label{eq:wedge_cov}
\end{equation}
with both index sums running over $\ell=0$, 2, 4, 6.  The correlation coefficient between the two wedges is then $r_{\perp\parallel}(k)=C_{\perp\parallel}/(C_{\perp\perp}C_{\parallel\parallel})^{1/2}$, as usual.  If the sums were continued to infinity, the two wedges would be uncorrelated under our approximation. Truncating the multipole sum at $\ell_{1,2}\le 6$ and assuming a linear theory $P_{\ell}$ gives $r_{\perp\parallel}\approx0.06$ for $\beta=f/b\approx 0.4$.  Sharp truncation in $\ell$ leads to ``leakage'' across $\mu_c$, with the leakage decreasing as further multipoles are included\footnote{The error bars in Fig.~\ref{fig:anisotropy} use the windowed Gaussian covariance of \S\ref{sec:analytical_fits}, truncated at $\ell=4$. The wedge auto-covariance (i.e.~errorbars in Fig.~\ref{fig:anisotropy}) is only marginally  ($\lesssim5\%$) affected by $\ell=6$. We have further confirmed that the two covariances predict similar results up to $\ell=4$, aside from large scales ($k\lesssim0.1$) that are significantly influenced by window matrices.}. On small scales, where the non-Gaussian covariance becomes important, the wedge-wedge correlation is expected to be larger.

The right panel of Figure~\ref{fig:anisotropy} compares the measured $P_{\perp}$ and $P_{\parallel}$ against the best-fit EFT model (convolved with the survey window function, \S\ref{sec:measurement}) obtained from the joint LAE\,$\times$\,QSO + QSO auto-spectrum fit. At large scales, both components are well described by the linear model, confirming that the anisotropic structure of the power spectrum is consistent with linear RSD and that no significant systematic contamination is present in either the transverse or radial directions. At small scales ($k \gtrsim 0.3\,h\,\mathrm{Mpc}^{-1}$) $P_{\perp}$ and $P_{\parallel}$ converge towards the same value, but $P_\perp < P_\parallel$ even at $k\approx 0.5 \,h\,\mathrm{Mpc}^{-1}$. This is similar to a linear theory prediction with no FoG, where supercluster infall enhances $P_{\parallel}$ above $P_\perp$ but both asymptote to shot noise at high $k$.  This model-agnostically indicates that the FoG suppression is rather weak and further indicates that there are no significant redshift errors, which would cause a damping of $P_\parallel$ compared to $P_\perp$.  This is qualitatively in agreement with past LAE measurements  \cite{Verhamme18,Uzsoy25}, which imply modest corrections ($k_{\parallel}\sigma\lesssim 1$). However, both studies emphasize the galaxy-population dependence of the redshift error, so a direct quantitative comparison with the current work is not straightforward.

The zero-crossing of the quadrupole is another important indicator of the strength of FoG effect \cite{BaleatoLizancos25}. From the measurement in Fig.~\ref{fig:multipoles}, we find that the zero-crossing is at high-$k$ ($k\gtrsim0.45\,h\,\mathrm{Mpc}^{-1}$) near the extent of the perturbative modeling regime ($k_{\rm nl}$) at $z\sim3$ \cite{Sailer21}. This indicates that the FoG suppression is weak, in agreement with the results observed through the wedge power spectra. 

\subsection{Fit to power spectrum}
\label{sec:FoG_fit}

The FoG effects for the power spectrum can be modeled using the EFT formalism up to one-loop order (\S\ref{sec:ps_theory}). In \S\ref{sec:analytical_fits} we mainly discuss the consequences of a Fourier-space fit at large scales. Here we focus on the FoG term $N_2$, in order to gain intuition of the strength of FoG effects for LAEs. 
The EFT fits to data indicate $N_2 = 4665 \pm 8445$ and $-1755\pm6630$ for the $z=2.5$ and 3.0 bins, respectively, in units of $h^{-5}\,\mathrm{Mpc}^5$. In neither case do we manage to detect the FoG effect at a statistically significant level.

\begin{figure}
    \centering
    \includegraphics[width=\textwidth]{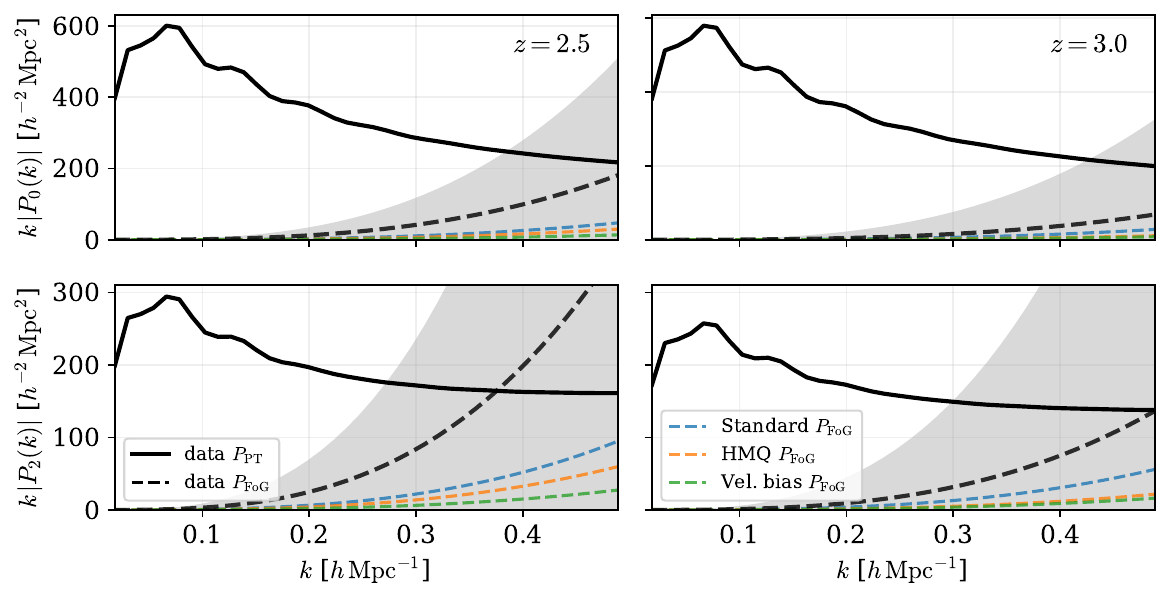}
    \caption{The FoG predictions from data (\S\ref{sec:analytical_fits}) and simulations (\S\ref{sec:HOD}), against the perturbation theory contribution $P_\mathrm{PT}$ fit from data. The columns are the $z$-bins 2.5 and 3.0 and the rows show the monopole $P_0$ and quadrupole $P_2$. The black solid and dashed lines show the $P_\mathrm{PT}$ and FoG from EFT fits to data, while the colored dashed lines show the FoG predictions from three best-fit HOD models. We observe that the the FoG predictions between data and HODs are in agreement, with both predicting relatively weak FoG signals. }
    \label{fig:pk_FoG}
\end{figure}

The best-fit HOD models from the HOD fits (\S\ref{sec:HOD}) also inform us of the FoG strength. 
For each model we measure the galaxy auto-spectrum monopole and quadrupole ($P^{gg}_0$, $P^{gg}_2$), real-space galaxy-matter cross-spectrum ($P^{gm}$), and real-space matter power spectrum ($P^{mm}$) over the periodic simulation box of 1$h^{-1}\,\mathrm{Gpc}$. For the galaxy auto-spectrum and cross-spectrum we use the Zeldovich Control Variates (ZCV) feature of the \textsc{AbacusUtils}, which reduces the sample variance on large scales by using Zeldovich predictions from the simulation initial conditions \cite{Kokron22,DeRose23,Hadzhiyska23,Bartlett26}. By jointly fitting these spectra with the same multi-tracer formalism as described in \S\ref{sec:analytical_fits}, we obtain measurements of the FoG. 

We compare the perturbation theory signal ($P_\mathrm{PT}$; the bias contributions) from data against the data- and simulation-derived FoG predictions in Fig.~\ref{fig:pk_FoG}. The all FoG predictions show relatively small contributions, with $P_2$ only becoming FoG dominated near $k\sim 0.3\,h\,\mathrm{Mpc}^{-1}$ at the upper end of the 1$\sigma$ uncertainty from data, and at even smaller $k\sim0.4\,h\,\mathrm{Mpc}^{-1}$ for $P_0$. Following the best-fit FoG from data, the FoG is even further suppressed, with only the $z=2.5$ $P_2$ becoming FoG dominated near $k\sim0.4\,h\,\mathrm{Mpc}^{-1}$ and the other spectra maintaining $P_\mathrm{PT}>|P_\mathrm{FoG}|$ to $k=0.5\,h\,\mathrm{Mpc}^{-1}\sim k_{\rm nl}$.

It is also notable that the HOD predictions for FoG are in agreement with the $1\sigma$ predictions from data, but generally predict weaker FoG. This is not a surprise, as the FoG estimates from the simulations likely underestimates, as typically $N_2$ scales with the shot noise, but we lack the mechanism to properly downsample the simulations. 

\section{Conclusion} \label{sec:conclusion}

In this work we analyze medium-band-selected Lyman alpha emitters (LAE), with the aim of characterizing these galaxies for next-generation, high-redshift cosmology surveys, such as DESI-II and Spec-S5 \cite{Schlegel22,Beseuner25}. We perform these analyses using LAE targets selected from the medium-band imaging data of the Intermediate Band Imaging Survey (IBIS) \cite{IBIS} and broadband data from the HSC-SSP wide field data \cite{HSC-SSP} which allows for high purity selection of LAEs. The DESI \texttt{tertiary54} ancillary program then spectroscopically follows up the targets with a high fiber completeness (of 98.6\%), obtaining a fiber-assignment-effect-free clustering sample with $\sim4000$ LAEs over a $\sim 6\deg^2$ footprint in the COSMOS field. The LAEs span a redshift range of $2.26<z<3.41$, which we divide into two redshift bins, centered at $z\approx2.5$ and 3.0.

We perform the analysis using the 3D correlation monopole and quadrupole, both in configuration and Fourier space. We fit the configuration measurements with a large grid of HOD simulations (\S\ref{sec:HOD}) from \texttt{AbacusUtils} \cite{Yuan22} which allows modeling of small (near one-halo) scales not describable by perturbative models. We perform the modeling using three variations of the HOD model, the standard five-parameter model \cite{Zheng07}, the HMQ model \cite{Alam20}, and the velocity bias model \cite{Guo15} and identify the best-fit parameter set for each framework using a covariance matrix constructed from 256 pseudo-independent mock catalogs matching the angular and radial distribution of data. The results indicate a general preference for the standard model, due to it  having the fewest parameters, and predict a large-scale bias of $b\sim2.2$ and 2.55 for the $z=2.5$ and 3.0 $z$-bins, respectively. 
Aside from the clustering modeling alone, the best-fit mock catalogs obtained here provide a ground for systematic tests. One particularly important to this work is confirming the independence of the sample from fiber assignment effects, an important artifact of DESI spectroscopy observations, as demonstrated in Appendix \ref{app:fiberassign}. 

The Fourier-space fits use the cross-correlation of LAEs and DESI DR2 QSOs \cite{DESI-DR2}, modeled using the EFT formalism standard in power spectrum analyses (\S\ref{sec:analytical_fits}). Taking advantage of the high-$z$ we fit the data to $k_\mathrm{max}=0.3 \,h\,\mathrm{Mpc}^{-1}$ with the one-loop power spectrum, with the Gaussian covariance matrix computed through \texttt{jaxpower} \cite{Li19}. The results provide $\approx10\%$ constraints on the linear bias: $b_1=2.31\pm0.23$ and $2.62\pm 0.20$ at $z=2.5$ and 3.0, respectively. These are the tightest constraints on the linear bias of LAEs to date, and provide key inputs into the cosmological information expected from next-generation surveys. One key quantity predicted from this is the power spectrum SNR at linear scales, which is $\bar{n}P_0 = 0.87$ and 0.85 at $k=0.1\,h\,\mathrm{Mpc}^{-1}$ for each $z$-bin. 

In addition to the standard EFT analysis, we explore the radiative transfer (RT) effect, capable of distorting the LAE clustering signal through complex selection effect induced by the scattering of Ly$\alpha$ photons off of \textsc{HI} clouds (\S\ref{sec:RT}). Given the current state of the data, we choose to model this with a very simple, leading order model involving a one-parameter ($\alpha$) distortion in the linear RSD signal \cite{Zheng11}.  We define $\alpha$ such that no RT effect corresponds to $\alpha=0$. The joint EFT-RT fits allows us to place first constraints on the strength of RT at $\alpha=-0.64\pm 0.63$ and $-0.95\pm0.55$ at $z=2.5$ and 3.0 bins, respectively. The results allow an order unity distortion, but are consistent with there being no RT signal to within $2\,\sigma$. 

Finally, we combine the measurements, HOD fits, and EFT fits to explore the strength of the FoG effect for LAEs (\S\ref{sec:FoG}). Specifically, we qualitatively observe the anisotropic distortion in $\xi(r_p,\pi)$, suppression in the line-of-sight power spectrum, and FoG fits to the data and HOD simulations. Each one of these probes suggest a relatively weak FoG contribution, which in turn suggests the cosmological information at small scales is still accessible. 

Although this work provides the best picture on the clustering of medium-band selected LAEs so far (building on previous work, e.g.~refs.~\cite{Ebina25,Raichoor25}), it is important that this effort continues in the future for theoretical readiness of next-generation cosmology. Using more physically motivated models, over the HOD models in this work, is one of the clear directions, as HOD models at best mimic the underlying physics of galaxy formation. This, for example, was evident particularly in the Poisson downsampling of mocks in this work, which does not mirror the true sampling of galaxies through imaging data and spectra emission and absorption. Another direction of possible interest is to explore more complex RT models. The one-parameter RT transfer model is not unreasonable, but past studies \cite{Wyithe11,Greig13} have predicted various other complex functional forms and these models have not been ruled out. Larger datasets or cross-correlation with other high-$z$ probes may provide enough statistical power to constrain a subset of these models with order unity or better. 

\section{Data Availability}

Material necessary to reproduce the figures in this work are publicly available at \url{https://doi.org/10.5281/zenodo.20750223}.

\section*{Acknowledgements}

The authors thank Edmond Chaussidon for useful discussions and for helpful comments on an earlier draft of the paper. HE and MW were supported by the DOE. 
This work made use of the Cobaya analysis code \cite{Cobaya,CobayaCode}, the jaxpower power spectra and window measurement code \cite{Li19}, the TheCoV covariance calculation code \cite{Wadekar19,Kobayashi23,Alves24}, the Abacus N-body simulation suite and AbacusUtils software \cite{Yuan22,Maksimova21,Garrison21}, the DESI redshift fitting pipeline RedRock \cite{Redrock}, and the correlation function measurement code Corrfunc \cite{Sinha20}.
This research was supported in part by grant NSF PHY-2309135 to the Kavli Institute for Theoretical Physics (KITP) and by the Aspen Center for Physics, which is supported by National Science Foundation grant PHY-2210452.

This material is based upon work supported by the U.S. Department of Energy (DOE), Office of Science, Office of High-Energy Physics, under Contract No. DE–AC02–05CH11231, and by the National Energy Research Scientific Computing Center, a DOE Office of Science User Facility under the same contract.
This material is based upon work supported by the U.S. Department of Energy (DOE), Office of Science, Office of High-Energy Physics, under Contract No. DE–AC02–05CH11231, and by the National Energy Research Scientific Computing Center, a DOE Office of Science User Facility under the same contract. Additional support for DESI was provided by the U.S. National Science Foundation (NSF), Division of Astronomical Sciences under Contract No. AST-0950945 to the NSF’s National Optical-Infrared Astronomy Research Laboratory; the Science and Technology Facilities Council of the United Kingdom; the Gordon and Betty Moore Foundation; the Heising-Simons Foundation; the French Alternative Energies and Atomic Energy Commission (CEA); the National Council of Humanities, Science and Technology of Mexico (CONAHCYT); the Ministry of Science, Innovation and Universities of Spain (MICIU/AEI/10.13039/501100011033), and by the DESI Member Institutions: \url{https://www.desi.lbl.gov/collaborating-institutions}. Any opinions, findings, and conclusions or recommendations expressed in this material are those of the author(s) and do not necessarily reflect the views of the U. S. National Science Foundation, the U. S. Department of Energy, or any of the listed funding agencies.
The authors are honored to be permitted to conduct scientific research on I'oligam Du'ag (Kitt Peak), a mountain with particular significance to the Tohono O’odham Nation.

This research draws upon DECam data as distributed by the Astro Data Archive at NSF NOIRLab. NOIRLab is managed by the Association of Universities for Research in Astronomy (AURA) under a cooperative agreement with the U.S. National Science Foundation.

This project used data obtained with the Dark Energy Camera (DECam), which was constructed by the Dark Energy Survey (DES) collaboration. Funding for the DES Projects has been provided by the US Department of Energy, the US National Science Foundation, the Ministry of Science and Education of Spain, the Science and Technology Facilities Council of the United Kingdom, the Higher Education Funding Council for England, the National Center for Supercomputing Applications at the University of Illinois at Urbana-Champaign, the Kavli Institute for Cosmological Physics at the University of Chicago, Center for Cosmology and Astro-Particle Physics at the Ohio State University, the Mitchell Institute for Fundamental Physics and Astronomy at Texas A\&M University, Financiadora de Estudos e Projetos, Fundação Carlos Chagas Filho de Amparo à Pesquisa do Estado do Rio de Janeiro, Conselho Nacional de Desenvolvimento Científico e Tecnológico and the Ministério da Ciência, Tecnologia e Inovação, the Deutsche Forschungsgemeinschaft and the Collaborating Institutions in the Dark Energy Survey.

The Collaborating Institutions are Argonne National Laboratory, the University of California at Santa Cruz, the University of Cambridge, Centro de Investigaciones Enérgeticas, Medioambientales y Tecnológicas–Madrid, the University of Chicago, University College London, the DES-Brazil Consortium, the University of Edinburgh, the Eidgenössische Technische Hochschule (ETH) Zürich, Fermi National Accelerator Laboratory, the University of Illinois at Urbana-Champaign, the Institut de Ciències de l’Espai (IEEC/CSIC), the Institut de Física d’Altes Energies, Lawrence Berkeley National Laboratory, the Ludwig-Maximilians Universität München and the associated Excellence Cluster Universe, the University of Michigan, NSF NOIRLab, the University of Nottingham, the Ohio State University, the OzDES Membership Consortium, the University of Pennsylvania, the University of Portsmouth, SLAC National Accelerator Laboratory, Stanford University, the University of Sussex, and Texas A\&M University.

Based on observations at NSF Cerro Tololo Inter-American Observatory, a program of NOIRLab (NOIRLab Prop. ID 2023B-184194 and 2025B-479281; PI: A.~Dey and D.~Schlegel), which is managed by the Association of Universities for Research in Astronomy (AURA) under a cooperative agreement with the U.S. National Science Foundation.

\appendix

\section{Fiber assignment}
\label{app:fiberassign}

To measure the clustering to small scales with spectroscopic datasets from DESI, it is necessary to account for the fiber assignment effects on clustering. The DESI instrumentation is limited such that any single observation cannot place fibers within $0.05\deg$ of one another, resulting in a preferential downselection of galaxy pairs nearby. At high-$z$, this issue is further amplified with the fiber collision scale $0.05\deg$ increasing up to $s\sim4\,h^{-1}\,\mathrm{Mpc}$ at $z\sim3$. While mitigation techniques for wide-field cosmology analyses have been developed \cite{Hawkins03,Anderson12,Reid14,Pinon25} they focus on regularizing the large scales important for cosmology and do not recover the lost information at small scales. 

As such, prior to making the \texttt{tertiary54} observations we studied what fiber assignment completeness was necessary in order that this systematic not unduly affect our observations.  We used mock catalogs developed from the best-fit HOD models in ref.~\cite{Ebina25}, generating 256 pseudo-independent catalogs (as done in \S\ref{sec:HOD}) using the $z$-distribution derived from applying the same color cuts to the \texttt{tertiary49} sample. Then, on each catalog, we ran fiber assignment using the same algorithm as used on the DESI data. From the pre- and post-fiber assignment catalogs, we each computed $\xi_\ell$ and its covariance. 

A similar test in ref.~\cite{Ebina25} found that, with their fiber assignment completeness ($\sim50\%$), the fiber assignment effects would inhibit the measurement of 3D clustering. Using the target selection and redshift distribution predictions from \texttt{tertiary49} we follow up these tests and identify that a very high fiber completeness would enable the first 3D clustering study of LAEs with no significant impact from fiber assignment effects. In particular, we find that for the sample in this work\footnote{This test is done with slightly larger footprint and hence lower fiber completeness, reflected in the small increase in $\chi^2$ as compared to the test after data collection.} the fiber assignment would influence measurements at a negligible $\chi^2<0.3$, including all scales below $4\,h^{-1}\,\mathrm{Mpc}$. This motivated the collection of high completeness data in \texttt{tertiary54}. 

\begin{figure}
    \centering
    \includegraphics[width=\linewidth]{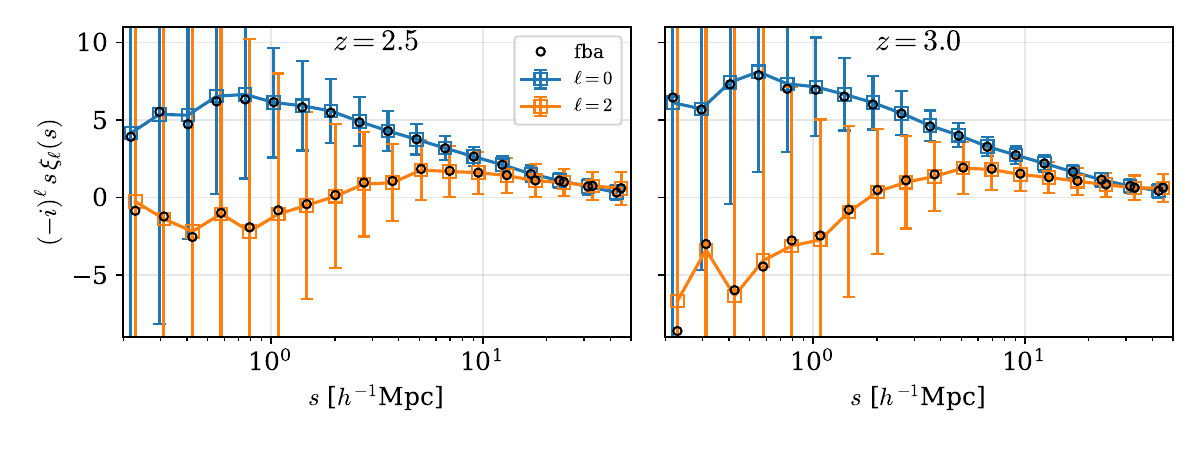}
    \caption{The comparison of mock correlation function measurements before (color) and after (black) fiber assignment (`fba'). The errors shown are calculated from the post-fiber assignment catalogs and predicts $\chi^2<0.1$ for each redshift, on fiber collision scales $s\lesssim 4\,h^{-1}\,\mathrm{Mpc}$.}
    \label{fig:xi_fba}
\end{figure}

After observation and clustering analyses, we again verify that this remains a negligible effect, using the best-fit standard HOD model derived in \S\ref{sec:HOD}. The pre- and post-fiber assignment clustering comparison, using the covariance computed from the post-fiber assignment clustering, is shown in Fig.~\ref{fig:xi_fba}. The results indicate $\chi^2<0.1$ below fiber collision scales, demonstrating that the errors from fiber assignment are completely negligible.

\section{Radial distribution of randoms}
\label{app:nz}

The radial distribution $n(z)$ assigned to the random catalog enters the density contrast through the ratio of data to random number densities. Any mismatch between the true selection function and the assumed $n(z)$ of the randoms either adds or subtracts large-scale power along the line of sight. Here we compare our default prescription to an alternative smoothing and quantify the impact on the measured power spectrum multipoles.

\begin{figure}
    \centering
    \includegraphics[width=\textwidth]{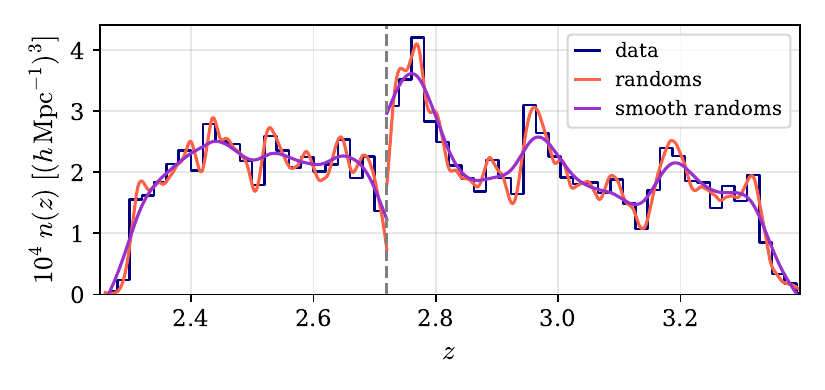}
    \caption{Comparison of the radial distribution $n(z)$ assigned to the random catalog using the default spline fit (red) and a smoother spline (purple), for the $z=2.5$ (left) and $z=3$ (right) redshift bins.}
    \label{fig:nz_comparison}
\end{figure}

As described in \S\ref{sec:clustering}, the default random catalog assigns radial positions by drawing redshifts from a smooth cubic spline fit to the observed $n(z)$ of the data, binned in intervals of $\Delta z = 0.005$. 
Fig.~\ref{fig:nz_comparison} compares this prescription to one with an order-of-magnitude larger smoothing for both redshift bins.  The smooth spline is clearly unable to capture all features of the selection function, e.g.~the sharp peaks at $z\sim 2.8$, 3.0, and 3.2, that recur with the same cadence as the medium bands (which have $\Delta z\sim 0.2$).

\begin{figure}
    \centering
    \includegraphics[width=\linewidth]{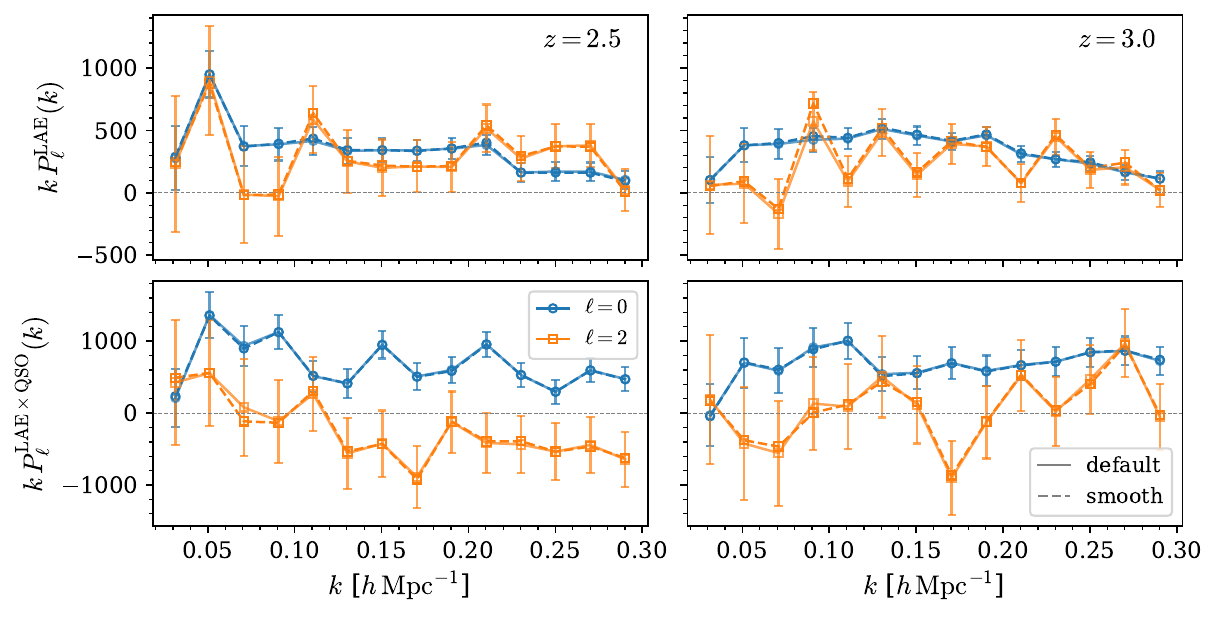}
    \caption{Comparison of the LAE auto-spectrum and LAE$\times$QSO cross-spectrum measurements using the default and smooth random prescriptions shown in Fig.~\ref{fig:nz_comparison}. The results indicate that the shifts are subdominant to the statistical errors already accounted for in the analysis. }
    \label{fig:pk_comparison}
\end{figure}

To assess the sensitivity of the clustering measurements to the choice of $n(z)$, we remeasure the LAE auto-power spectrum multipoles $P_0(k)$ and $P_2(k)$ using the two prescriptions for selecting redshifts for the randoms, while keeping all other aspects of the analysis (FKP weights, window function, covariance) identical. Fig.~\ref{fig:pk_comparison} shows the resulting power spectra, demonstrating that while there are visually identifiable differences, the largest differences come in the quadrupole, where the error bars are sufficiently large to make the effect subdominant to statistical errors. 
Nonetheless, this additional, subdominant uncertainty is a variation of the radial integral constraint and stems from the small survey area of this sample. Future studies with larger area datasets will be less susceptible to this issue and hence will be able to confirm the measurements here with higher precision. 


\section{Author Affiliations}
\label{sec:affiliations}

\noindent \hangindent=.5cm $^{a}${Department of Physics, University of California, Berkeley, CA 94720, USA}

\noindent \hangindent=.5cm $^{b}${Berkeley Center for Cosmological Physics, UC Berkeley, CA 94720, USA}

\noindent \hangindent=.5cm $^{c}${Lawrence Berkeley National Laboratory, 1 Cyclotron Road, Berkeley, CA 94720, USA}

\noindent \hangindent=.5cm $^{d}${NSF NOIRLab, 950 N. Cherry Ave., Tucson, AZ 85719, USA}

\noindent \hangindent=.5cm $^{e}${Department of Physics, Boston University, 590 Commonwealth Avenue, Boston, MA 02215 USA}

\noindent \hangindent=.5cm $^{f}${Dipartimento di Fisica ``Aldo Pontremoli'', Universit\`a degli Studi di Milano, Via Celoria 16, I-20133 Milano, Italy}

\noindent \hangindent=.5cm $^{g}${INAF-Osservatorio Astronomico di Brera, Via Brera 28, 20122 Milano, Italy}

\noindent \hangindent=.5cm $^{h}${Department of Physics \& Astronomy, University College London, Gower Street, London, WC1E 6BT, UK}

\noindent \hangindent=.5cm $^{i}${Institut d'Estudis Espacials de Catalunya (IEEC), c/ Esteve Terradas 1, Edifici RDIT, Campus PMT-UPC, 08860 Castelldefels, Spain}

\noindent \hangindent=.5cm $^{j}${Institute of Space Sciences, ICE-CSIC, Campus UAB, Carrer de Can Magrans s/n, 08913 Bellaterra, Barcelona, Spain}

\noindent \hangindent=.5cm $^{k}${Department of Physics and Astronomy, The University of Utah, 115 South 1400 East, Salt Lake City, UT 84112, USA}

\noindent \hangindent=.5cm $^{l}${Instituto de F\'{\i}sica, Universidad Nacional Aut\'{o}noma de M\'{e}xico,  Circuito de la Investigaci\'{o}n Cient\'{\i}fica, Ciudad Universitaria, Cd. de M\'{e}xico  C.~P.~04510,  M\'{e}xico}

\noindent \hangindent=.5cm $^{m}${University of California, Berkeley, 110 Sproul Hall \#5800 Berkeley, CA 94720, USA}

\noindent \hangindent=.5cm $^{n}${Instituci\'{o} Catalana de Recerca i Estudis Avan\c{c}ats, Passeig de Llu\'{\i}s Companys, 23, 08010 Barcelona, Spain}

\noindent \hangindent=.5cm $^{o}${Institut de F\'{i}sica d’Altes Energies (IFAE), The Barcelona Institute of Science and Technology, Edifici Cn, Campus UAB, 08193, Bellaterra (Barcelona), Spain}

\noindent \hangindent=.5cm $^{p}${Departamento de F\'isica, Universidad de los Andes, Cra. 1 No. 18A-10, Edificio Ip, CP 111711, Bogot\'a, Colombia}

\noindent \hangindent=.5cm $^{q}${Observatorio Astron\'omico, Universidad de los Andes, Cra. 1 No. 18A-10, Edificio H, CP 111711 Bogot\'a, Colombia}

\noindent \hangindent=.5cm $^{r}${University of Virginia, Department of Astronomy, Charlottesville, VA 22904, USA}

\noindent \hangindent=.5cm $^{s}${Departamento de F\'{\i}sica, DCI-Campus Le\'{o}n, Universidad de Guanajuato, Loma del Bosque 103, Le\'{o}n, Guanajuato C.~P.~37150, M\'{e}xico}

\noindent \hangindent=.5cm $^{t}${Fermi National Accelerator Laboratory, PO Box 500, Batavia, IL 60510, USA}

\noindent \hangindent=.5cm $^{u}${Department of Astronomy, University of Texas at Austin, 2515 Speedway, TX 78712, USA}

\noindent \hangindent=.5cm $^{v}${Institut d'Astrophysique de Paris. 98 bis boulevard Arago. 75014 Paris, France}

\noindent \hangindent=.5cm $^{w}${IRFU, CEA, Universit\'{e} Paris-Saclay, F-91191 Gif-sur-Yvette, France}

\noindent \hangindent=.5cm $^{x}${Department of Physics, The University of Texas at Dallas, 800 W. Campbell Rd., Richardson, TX 75080, USA}

\noindent \hangindent=.5cm $^{y}${Department of Physics and Astronomy, University of California, Irvine, 92697, USA}

\noindent \hangindent=.5cm $^{z}${Perimeter Institute for Theoretical Physics, 31 Caroline St. North, Waterloo, ON N2L 2Y5, Canada}

\noindent \hangindent=.5cm $^{aa}${Sorbonne Universit\'{e}, CNRS/IN2P3, Laboratoire de Physique Nucl\'{e}aire et de Hautes Energies (LPNHE), FR-75005 Paris, France}

\noindent \hangindent=.5cm $^{ab}${Departament de F\'{i}sica, Serra H\'{u}nter, Universitat Aut\`{o}noma de Barcelona, 08193 Bellaterra (Barcelona), Spain}

\noindent \hangindent=.5cm $^{ac}${Center for Cosmology and AstroParticle Physics, The Ohio State University, 191 West Woodruff Avenue, Columbus, OH 43210, USA}

\noindent \hangindent=.5cm $^{ad}${Department of Astronomy, The Ohio State University, 4055 McPherson Laboratory, 140 W 18th Avenue, Columbus, OH 43210, USA}

\noindent \hangindent=.5cm $^{ae}${The Ohio State University, Columbus, 43210 OH, USA}

\noindent \hangindent=.5cm $^{af}${Institute of Cosmology and Gravitation, University of Portsmouth, Dennis Sciama Building, Portsmouth, PO1 3FX, UK}

\noindent \hangindent=.5cm $^{ag}${Department of Physics and Astronomy, University of Waterloo, 200 University Ave W, Waterloo, ON N2L 3G1, Canada}

\noindent \hangindent=.5cm $^{ah}${Waterloo Centre for Astrophysics, University of Waterloo, 200 University Ave W, Waterloo, ON N2L 3G1, Canada}

\noindent \hangindent=.5cm $^{ai}${Instituto de Astrof\'{i}sica de Andaluc\'{i}a (CSIC), Glorieta de la Astronom\'{i}a, s/n, E-18008 Granada, Spain}

\noindent \hangindent=.5cm $^{aj}${Departament de F\'isica, EEBE, Universitat Polit\`ecnica de Catalunya, c/Eduard Maristany 10, 08930 Barcelona, Spain}

\noindent \hangindent=.5cm $^{ak}${Department of Physics and Astronomy, Sejong University, 209 Neungdong-ro, Gwangjin-gu, Seoul 05006, Republic of Korea}

\noindent \hangindent=.5cm $^{al}${CIEMAT, Avenida Complutense 40, E-28040 Madrid, Spain}

\noindent \hangindent=.5cm $^{am}${Department of Physics, University of Michigan, 450 Church Street, Ann Arbor, MI 48109, USA}

\noindent \hangindent=.5cm $^{an}${University of Michigan, 500 S. State Street, Ann Arbor, MI 48109, USA}

\noindent \hangindent=.5cm $^{ao}${Department of Physics \& Astronomy, Ohio University, 139 University Terrace, Athens, OH 45701, USA}

\bibliographystyle{JHEP}
\bibliography{main}

\end{document}